\documentclass[pre,twocolumn,showpacs]{revtex4}
\usepackage{graphicx}
\usepackage{amsfonts}

\newcommand{\ba}{\begin{eqnarray}}
\newcommand{\be}{\begin{equation}}
\newcommand{\ea}{\end{eqnarray}}
\newcommand{\ee}{\end{equation}}

\newcommand{\ignore}[1]{}

\begin{document}

\title{Random transition-rate matrices for the master equation}

\author{Carsten Timm}
\email{carsten.timm@tu-dresden.de}
\affiliation{Institute for Theoretical Physics, Technische Universit\"at
Dresden, 01062 Dresden, Germany}

\date{May 16, 2000}

\begin{abstract}
Random-matrix theory is applied to transition-rate matrices in the Pauli master
equation. We study the distribution and correlations of
eigenvalues, which govern the dynamics of complex stochastic systems. Both
the cases of identical and of independent rates of forward and backward
transitions are considered. The first case leads to symmetric transition-rate
matrices, whereas the second corresponds to general, asymmetric matrices. The
resulting matrix ensembles are different from the standard ensembles and
show different eigenvalue distributions. For example, the fraction of real
eigenvalues scales anomalously with matrix dimension in the asymmetric case.
\end{abstract}


\maketitle

\section{Introduction}
\label{sec.intro}

The Pauli master equation is encountered in many fields of science such as
physics, chemistry, and biology. It describes the time evolution of
probabilites for a system to be in certain states. Formally identical
rate equations describe the dynamics of concentrations or populations of certain
entities. The dynamics of probabilities is described by the Pauli master
equation
\be
\dot P_i = \sum_{j\neq i} ( R_{ij}P_j - R_{ji}P_i ) ,
\label{1.rate}
\ee
where $P_i$ is the probability to find the system in state $i=1,\ldots,N$
and $R_{ij}$ is the transition rate from state $j$ to state $i$. Evidently, the
rates of change of probabilities depend only on the probabilities at time $t$,
i.e., Eq.~(\ref{1.rate}) describes a memory-less or Markovian process.
Equation (\ref{1.rate}) ensures that the total probability is conserved,
\be
\frac{d}{dt} \sum_i P_i = \sum_{ij,i\neq j} ( R_{ij}P_j - R_{ji}P_i ) = 0 .
\label{1.Pcons}
\ee
Typical applications in physics include lasers \cite{Sie86}, disordered
conductors \cite{ABS81}, microelectronic devices \cite{Car02}, quantum dots
\cite{GlM88}, and molecular electronics \cite{GRN07}. In these cases one can, in
principle, obtain the Pauli master equation by first deriving a \emph{quantum}
master equation
for the reduced density matrix of a small system, which is obtained by tracing
out the reservoir degrees of freedom from the full density operator
\cite{WaB53,Red65,ToM75,BrP02,Tim08}. If the off-diagonal
components of the reduced density matrix decay rapidly, it is
sufficient to keep only the diagonal components representing the probabilities
$P_i$ of states $|i\rangle$ of the small system. In certain fields, for example
in transport and laser theory, the resulting Eqs.~(\ref{1.rate}) are often
called \emph{rate equations}.

However, if even the small system is complicated, such as a system of
interacting enzymes, this route becomes
unfeasible. In applications outside of physics, where $i$ could refer to the
state of a technical or social process, a quantum-statistical description
becomes inappropriate in any case.
One would then view Eq.~(\ref{1.rate}) as the fundamental description.

Our goal is to make progress in the understanding of the master
equations for complex systems. The number $N$ of possible states will typically
be large. It should be noted however that complex behavior can already emerge
for moderate $N$. An example is provided by the differential conductance
calculated in Ref.~\cite{Tim07} for a magnetic molecule with magnetic anisotropy
axis not aligned with the applied magnetic field, where $N=20$, but
due to noncommuting terms in the Hamiltonian many rates are nonzero and are
distributed over a broad range.

\subsection{Properties of the master equation}
\label{sub.prop}

We first recount some basic properties. It is clear that one can rewrite
Eq.~(\ref{1.rate}) in the form
\be
\dot P_i = \sum_j A_{ij} P_j
\label{1.stocheq}
\ee
or $\dot\mathbf{P} = A\mathbf{P}$ with the transition-rate matrix, or,
for short, rate matrix,
\be
A_{ij} \equiv \left\{\begin{array}{ll}
  R_{ij} & \mbox{for $i\neq j$} \\[1ex]
  -\sum_{k\neq j} R_{kj} & \mbox{for $i=j$.}
  \end{array}\right.
\label{1.Adef}
\ee
It follows that the column sums vanish,
\be
\sum_i A_{ij} = 0 \quad\mbox{for all $j$.}
\label{1.Aconstr}
\ee
Note that $(d/dt)\, \sum_i P_i = \sum_{ij} A_{ij} P_j$ vanishes for all $P_j$
if and only if Eq.~(\ref{1.Aconstr}) holds. The constraint
(\ref{1.Aconstr}) is thus dictated by conservation of probability. From
Eq.~(\ref{1.Adef}) it is also clear that
\be
A_{ij} \ge 0 \quad\mbox{for all $i\neq j$}
\label{1.Apos}
\ee
if we interpret the $R_{ij}$ as transition rates. A matrix satisfying the
inequalities (\ref{1.Apos}) and $\sum_i A_{ij} \le 0$ for all $j$ is called a
\emph{compartmental matrix}.


Equation (\ref{1.stocheq}) can be solved by the ansatz $\mathbf{P} = e^{\lambda
t} \mathbf{v}$, which leads to the eigenvalue equation
$A\mathbf{v}=\lambda\mathbf{v}$.
Since $A$ is generally not symmetric, the eigenvalues $\lambda$ and the
components of the right eigenvectors $\mathbf{v}$ can be
complex. However, since $A$ is real, the equation
$A\mathbf{v}=\lambda\mathbf{v}$ implies
$A\mathbf{v}^\ast=\lambda^\ast\mathbf{v}^\ast$. Thus, the eigenvalues are
real with real eigenvectors or form complex conjugate pairs with their
eigenvectors also being complex conjugates.

Let $\mathbf{v}_n$ be the right eigenvector to eigenvalue $\lambda_n$.
It is well known that
there is always at least one strictly zero eigenvalue, which we call
$\lambda_0=0$: the constraint (\ref{1.Aconstr}) implies that $A$ has a left
eigenvector $(1,1,\ldots,1)$ to the eigenvalue $\lambda_0=0$.
The corresponding right eigenvector $\mathbf{v}_0$ describes the stationary
state.


A real eigenvector $\mathbf{v}_n$ with real eigenvalue $\lambda_n$
describes a contribution to the probability vector $\mathbf{P}$ that decays
exponentially with the rate $-\lambda_n$. A complex conjugate pair of
eigenvectors $\mathbf{v}_n$, $\mathbf{v}_n^\ast$ with eigenvalues $\lambda_n$,
$\lambda_n^\ast$ can be combined to form the two independent real solutions
$(e^{\lambda_n t} \mathbf{v}_n + e^{\lambda_n^\ast t} \mathbf{v}_n^\ast)/2$ and
$(e^{\lambda_n t} \mathbf{v}_n - e^{\lambda_n^\ast t} \mathbf{v}_n^\ast)/2i$.
Writing the components of $\mathbf{v}_n$ as $v_{nj} = v_{nj}^0 e^{i\phi_{nj}}$
with $v_{nj}^0$ real, we obtain the solutions
\be
v_{nj}^0\, e^{\mathrm{Re}\,\lambda_n t} \, \times
  \left\{\begin{array}{l}
    \cos(\mathrm{Im}\,\lambda_n t+\phi_{nj}) \\[1ex]
    \sin(\mathrm{Im}\,\lambda_n t+\phi_{nj}) .
  \end{array}\right.
\label{1.dampos}
\ee
The initial values at time $t=0$ are clearly $\mathrm{Re}\,v_{nj}$ and
$\mathrm{Im}\,v_{nj}$, respectively.
We thus find damped harmonic oscillations with damping rate
$-\mathrm{Re}\,\lambda_n$ and angular frequency $\mathrm{Im}\,\lambda_n$.
We obtain the solution at all times by expanding the initial probability vector
$\mathbf{P}(t=0)$ into the basis of real vectors $\mathbf{v}_n$ (for real
$\lambda_n$) and $\mathrm{Re}\,\mathbf{v}_n$, $\mathrm{Im}\,\mathbf{v}_n$ (for
complex conjugate pairs $\lambda_n$, $\lambda_n^\ast$).


An eigenvalue $\lambda_n$ with $\mathrm{Re}\,\lambda_n>0$ would be unphysical,
since the corresponding contribution to the
probabilities would diverge for $t\to\infty$. However, for any
compartmental matrix the spectrum is contained in
$\{\lambda|\mathrm{Re}\,\lambda<0\} \cup \{0\}$ \cite{BeP79,Hof96}. Thus all
eigenvalues are either zero or have a strictly negative real part.


The Perron-Frobenius theorem \cite{Per07,Fro12} applied to the
non-negative matrix $A-a_\mathrm{min}I$, where $a_\mathrm{min}<0$ is the minimum
of $A_{ii}$ and $I$ is the $N\times N$ unit matrix, shows that the right
eigenvector $\mathbf{v}_0$ to $\lambda_0$ has
only non-negative components. This ensures that the probabilities in the
stationary state are non-negative.

\subsection{Random rate matrices}

As noted above, even relatively simple problems lead to master equations with
rates $A_{ij}$, $i\neq j$, distributed over a broad range. In problems with
large numbers of states it is often impractical to obtain all
independent components $A_{ij}$. This situation is reminiscent of Hamiltonians
for complex systems. Difficult problems of this type concern
atomic nuclei and quantum dots,
where the Hamiltonian is too complicated to write
down explicitly, but cannot be simplified by methods restricted to weakly
interacting systems. For these systems, random-matrix theory (RMT)
\cite{Wig57,Bee97,GMW98,Meh04} has lead
to significant progress. The main assumption is that a Hamiltonian of this
type is a typical representative of an ensemble of Hamiltonians of appropriate
symmetry. While this
approach does not allow one to obtain specific eigenvalues, it does provide
information about the statistical properties of the spectrum
\cite{Wig57,Bee97,GMW98,Meh04}.

Our point of departure is to treat the rate matrix
$A$ for a complex system as an element of a suitable
random-matrix ensemble. In the case of transport through quantum dots, this is
complementary to treating the Hamiltonian of the quantum dot
as a random matrix, which has been done extensively \cite{Bee97}.

Since the rate matrix $A$ must satisfy the conditions (\ref{1.Aconstr})
and (\ref{1.Apos}), we define the \emph{exponential
general rate-matrix ensemble} (EGRE): The EGRE is formed by real $N\times N$
matrices $A$ with independently identically distributed off-diagonal components
$A_{ij}$ with the distribution function
\be
p(A_{ij}) = \left\{\begin{array}{ll}
\displaystyle\frac{1}{\langle R\rangle}\,e^{-A_{ij}/\langle R\rangle} &
  \mbox{for $A_{ij}\ge 0$} \\[1.5ex]
0 & \mbox{otherwise}
\end{array}\right.
\label{1.expdist}
\ee
and the diagonal components
\be
A_{jj} = -\sum_{i\neq j} A_{ij} .
\label{1.Adiag}
\ee
The exponential distribution of rates $A_{ij}$ is viewed as the least biased
distribution of non-negative numbers. We will also present
results that do not depend on the specific distribution function $p$. We will
see that the
specific distribution becomes irrelevant in the limit of large $N$, at least if
all its moments exist. The distribution of components is thus not the
most fundamental difference between the EGRE and the well-known ensembles
studied in the context of random Hamiltonians. Rather, one such difference lies
in the constraint (\ref{1.Aconstr}) or (\ref{1.Adiag}). The other is that
the rate matrices are real but not symmetric and thus not hermitian
\cite{footnote.Amir}.

Ensembles of non-hermitian matrices have been studied in detail, starting with
Ginibre's work on Gaussian ensembles of non-hermitian matrices with real,
complex, and quaternion components \cite{Gin65}. We will compare our results to
the real Ginibre ensemble.

To be able to analyze the importance of the asymmetry,
we also define the \emph{exponential
symmetric rate-matrix ensemble} (ESRE): The ESRE is formed by real
symmetric $N\times N$ matrices $A$ with independently identically distributed
components $A_{ij}$ above the diagonal ($i<j$) with the distribution function
given by Eq.~(\ref{1.expdist})
and the diagonal components given by Eq.~(\ref{1.Adiag}).

Another possible choice is a two-valued distribution of rates, where a
transition from state $j$ to state $i$ is either possible or impossible, and all
possible transitions have the same rate. This case with symmetric rates
has been studied by various authors \cite{BrR88,Fyo99,SMF03}. It is
essentially equivalent to adjacency matrices of random simple networks.


An ensemble of real symmetric matrices satisfying Eq.~(\ref{1.Aconstr}) but with
a Gaussian distribution of $A_{ij}$ has also been studied \cite{SMF03}. This
case cannot easily be interpreted in terms of a master equation, since the
$A_{ij}$ can be negative. We will compare our results for the eigenvalue
spectrum to these works below.

The remainder of this paper is organized as follows: In Sec.~\ref{sec.ESRE} we
consider the simpler case of symmetric rate matrices (the ESRE) and obtain
results for the eigenvalue density and for the correlations between neighboring
eigenvalues. In Sec.~\ref{sec.EGRE} we then study general rate matrices
(the EGRE) and obtain results for the eigenvalue density, now in the complex
plane, and for the correlations of neighboring eigenvalues. We conclude in
Sec.~\ref{sec.conc}. A number of analytical derivations are relegated to
appendices.

\section{Symmetric rate-matrix ensemble}
\label{sec.ESRE}

We first consider ensembles of symmetric rate matrices $A$. These
describe processes where transitions from any state $j$ to state $i$ and from
$i$ to $j$ occur with the same rate, $A_{ij}=A_{ji}$.

\subsection{Spectrum}
\label{sus.ESRE.spec}

As noted above, the spectrum always contains the eigenvalue $\lambda_0=0$. The
corresponding
eigenvector for symmetric matrices is $(1,1,\ldots,1)$ or, normalized to unit
probability, $(1/N,1/N,\ldots,1/N)$. For symmetric rates, the
stationary state is thus characterized by equal distribution over all states
$i$. We are interested in the distribution of the other
eigenvalues $\lambda_n$, $n=1,\ldots, N-1$ , which are all real. We have also
seen in Sec.~\ref{sub.prop} that $\lambda_n\le 0$. Since there is no further
constraint, the probability of $\lambda_n$ for any $n>0$ being exactly zero
vanishes.

To simplify the calculations, we shift the matrices so that they have zero mean.
We discuss this immediately for general matrices.
Also, nothing here depends on the distribution function $p$ of the rates
$A_{ij}$, as long as the average $\langle R\rangle \equiv \langle A_{ij}\rangle$
exists. We define
\be
\tilde A \equiv A - \langle A\rangle ,
\label{2.shiftA}
\ee
where here and in the following angular brackets denote the average over the
matrix ensemble under consideration. Here, $\langle A\rangle$ has the components
$\langle A_{ij}\rangle = \langle R\rangle$ for $i\neq j$ and
$\langle A_{ii}\rangle = -(N-1)\,\langle R\rangle$.
Is follows that $\sum_i \tilde A_{ij} = 0$ for all $j$. Consequently, $\tilde
A$ has a left eigenvector $\mathbf{w}_0^T\equiv (1,1,\ldots,1)$ to the
eigenvalue $\tilde\lambda_0=0$.

Let $\mathbf{v}_n$ be the right eigenvectors of $A$ to the eigenvalues
$\lambda_n$, $n=1,\ldots,N-1$. Since $\mathbf{w}_0^T$ is the left eigenvector
to the eigenvalue $\lambda_0=0$, we have $\mathbf{w}_0^T \mathbf{v}_n=0$. Since
\be
\langle A\rangle = \langle R\rangle\, \left(\begin{array}{ccc}
  1 & \cdots & 1 \\
  \vdots & \ddots & \vdots \\
  1 & \cdots & 1
  \end{array}\right) - N \langle R\rangle\, I ,
\ee
$\mathbf{v}_n$ is a right eigenvector of $\langle A\rangle$ to the
eigenvalue $-N\langle R\rangle$. Therefore, $\mathbf{v}_n$ is also a
right eigenvector of $\tilde A$ to the eigenvalue $\tilde\lambda_n = \lambda_n
+ N\langle R\rangle$.
The result is that the shifted matrices $\tilde A$ also have one eigenvalue
$\tilde\lambda_0=0$ and that the remaining eigenvalues are just the eigenvalues
of $A$, shifted by $N\langle R\rangle$.

We now derive the average of eigenvalues $\lambda_n$, here and in the
following excluding $\lambda_0=0$. We have
$\langle \lambda \rangle' = \langle\tilde\lambda\rangle' - N\langle R\rangle$,
where angular brackets with a prime denote the average over all eigenvalues,
excluding the exact zero. Since this leaves $N-1$ eigenvalues, their average is
the trace of the matrix, to which the zero eigenvalue does not contribute,
divided by $N-1$. Consequently,
\be
\langle \tilde\lambda\rangle' = \frac{1}{N-1}\,\mathrm{Tr}\,\langle \tilde
A\rangle = \frac{1}{N-1}\,\mathrm{Tr}\,0 = 0
\ee
so that
\be
\langle \lambda \rangle' = -N\langle R\rangle .
\label{2.meanla}
\ee
This result is independent of the specific distribution function of rates, $p$,
as long as $\langle R\rangle$ exists.

We next calculate the low-order central moments
\be
\mu_m \equiv \langle \tilde\lambda^m \rangle'
  = \langle (\lambda -\langle\lambda\rangle')^m\rangle'
  = \langle (\lambda + N\langle R\rangle)^m\rangle'
\ee
of the eigenvalues $\lambda_n$, $n>0$. The central moments are
identical to the central moments of the shifted values $\tilde\lambda_n$. 
Unless otherwise noted, our results for $\mu_m$ hold for an arbitrary
distribution function of rates, $p$, as long as the moments exist.
It is instructive to show the calculation of the
second moment explicitly. We find
\ba
\mu_2 & = & \langle \tilde\lambda^2\rangle'
  = \frac{1}{N-1}\,\mathrm{Tr}\,\langle \tilde A^2\rangle
  = \frac{1}{N-1}\,\sum_{ij} \langle \tilde A_{ij} \tilde A_{ji} \rangle
  \nonumber \\
& = & \frac{1}{N-1}\, \sum_i \bigg( \sum_{j\neq i} \langle \tilde A_{ij}
  \tilde A_{ji} \rangle
  + \langle \tilde A_{ii} \tilde A_{ii}\rangle \bigg) .
\label{2.mu2a}
\ea
Using $\tilde A_{ij}=\tilde A_{ji}$ and $\sum_k \tilde A_{ki}=0$, we
obtain
\be
\mu_2 = \frac{1}{N-1}\, \sum_i \left( \sum_{j\neq i} \langle \tilde A_{ij}^2
\rangle
+ \sum_{k,l\neq i} \langle \tilde A_{ki} \tilde A_{li}\rangle \right) .
\ee
With $\langle\tilde A_{ij}\rangle=0$ we finally get
\be
\mu_2 = \frac{2}{N-1}\, \sum_i \sum_{j\neq i} \langle \delta R^2\rangle
  = 2N\,\langle \delta R^2\rangle ,
\label{2.mu2c}
\ee
where $\langle \delta R^2\rangle \equiv \langle A_{ij}^2 \rangle - \langle
A_{ij}\rangle^2$ for $i\neq j$ is the second central moment of $p(A_{ij})$. For
the special case of an exponential distribution we
have $\langle \delta R^2\rangle = \langle R\rangle^2$ and
thus $\mu_2 = 2N\,\langle R\rangle^2$.

The important consequence is that while the mean of the nonzero eigenvalues
of the unshifted matrices $A$ scales with $N$, Eq.~(\ref{2.meanla}), the width
of their distribution is only $\sqrt{\mu_2} = \sqrt{2N\, \langle \delta
R^2\rangle} \propto \sqrt{N}$. Thus for large $N$ the distribution of
eigenvalues contains a single eigenvalue $\lambda_0=0$ and the remaining
$N-1$ eigenvalues form a narrow distribution around $-N\langle R\rangle$.
In physical terms, nearly all deviations from the stationary state decay on the
same time scale $1/N\langle R\rangle$.

All moments can be obtained by the same method: We first write the
average in terms of a trace, split the sum into terms with equal or distinct
matrix indices, and use $\sum_k \tilde A_{ki}=0$. With
$\tilde A_{ij}=\tilde A_{ji}$ and $\langle \tilde A_{ij}\rangle=0$ we obtain the
moments. Since the enumeration of all possible cases of equal or
distinct indices is cumbersome, we have used a symbolic algebra scheme
implemented with Mathematica \cite{Mat08}. The results up to $m=8$ are
shown in Table \ref{tab.1} for a general distribution. The
moments are expressed in terms of the central moments $\langle \delta R^n\rangle
\equiv \langle (A_{ij}-\langle A_{ij}\rangle)^n\rangle$.
Note that in the limit of large $N$, the moments $\mu_m$ for
even $m$ only depend on the second moment $\langle \delta R^2\rangle$. We will
return to this point shortly.


\begin{table*}
\caption{\label{tab.1}Central moments $\mu_m$, $m=2,\ldots,8$, of the
nonzero eigenvalues $\lambda$ for ensembles of symmetric
rate matrices. The results hold
independently of the distribution function $p$ of rates $A_{ij}$, $i<j$, as long
as the moments exist. Here, $\langle\delta R^n\rangle$ is the $n$-th central
moment of $p$.}
\begin{ruledtabular}
\begin{tabular}{cc}
$m$ & $\mu_m$ (symmetric matrices, general distribution) \\
\hline
$2$ & $2N \langle \delta R^2\rangle$ \\
$3$ & $-4N \langle \delta R^3\rangle$ \\
$4$ & $N[9(N-2)\langle \delta R^2\rangle^2 + 8 \langle \delta R^4\rangle]$ \\
$5$ & $-2N [ 25(N-2) \langle \delta R^2\rangle\langle \delta R^3\rangle
  + 8\langle \delta R^5\rangle ]$ \\
$6$ & $N [4 (14N^2-73N+90)\langle \delta R^2\rangle^3 + 73(N-2)\langle\delta
  R^3\rangle^2 + 132 (N-2) \langle\delta R^2\rangle\langle \delta R^4\rangle
  + 32 \langle \delta R^6\rangle]$ \\
$7$ & $-2N [7 (41N^2-211N+258) \langle \delta R^2\rangle^2 \langle \delta
  R^3\rangle + 203(N-2) \langle \delta R^3\rangle\langle \delta R^4\rangle
  + 168 (N-2) \langle \delta R^2\rangle \langle \delta R^5\rangle
  + 32 \langle\delta R^7\rangle]$ \\
$8$ & $N [(431N^3 - 4042N^2 + 12021N -11322) \langle \delta R^2\rangle^4
  + 6 (306N^2 - 1561N + 1898) \langle \delta R^2\rangle^2 \langle \delta
  R^4\rangle + 593(N-2) \langle \delta R^4\rangle^2$ \\
 & ${}+ 1088(N-2) \langle \delta R^3\rangle \langle \delta R^5\rangle
  + 4 (N-2) (507N-1574) \langle\delta R^2\rangle\langle\delta R^3\rangle^2
  + 832 (N-2) \langle\delta R^2\rangle \langle\delta R^6\rangle
  + 128 \langle \delta R^8\rangle]$ \\
\end{tabular}
\end{ruledtabular}
\end{table*}

Table \ref{tab.2} shows the central moments $\mu_m$ up to $m=10$ for the
exponential distribution of $A_{ij}$, $i<j$ (ESRE). For the exponential
distribution, one has
$\langle \delta R^n\rangle = {}!n\,\langle R\rangle^n$, where $!n \equiv
n!\sum_{k=0}^n (-1)^k/k!$ is the subfactorial. Table \ref{tab.2} also contains
the leading large-$N$ terms for the ESRE. At least up to $m=10$, the even
moments scale as $\mu_m \sim N^{m/2}$ for large $N$, as
expected from the scaling of $\mu_2$. However, the odd
moments scale only as $\mu_m \sim N^{(m-1)/2}$. If this holds for all
$m$, the distribution of $\tilde\lambda$ approaches an even function
for large $N$. This is indeed the case, as we shall see.

\begin{table*}
\caption{\label{tab.2}Second column: central moments $\mu_m$, $m=2,\ldots,10$,
of the nonzero eigenvalues $\lambda$ for ensembles of symmetric rate
matrices, assuming an exponential distribution of rates (ESRE). Third column:
leading term of $\mu_m$ for large $N$.}
\begin{ruledtabular}
\begin{tabular}{ccc}
$m$ & $\mu_m$ (ESRE) & $\mu_m$ (ESRE, $N\gg 1$) \\
\hline
$2$ & $2N \langle R\rangle^2$ & $2N \langle R\rangle^2$ \\
$3$ & $-8N \langle R\rangle^3$ & $-8N \langle R\rangle^3$ \\
$4$ & $9N(N+6)\langle R\rangle^4$ & $9N^2 \langle R\rangle^4$ \\
$5$ & $-4N(25N+126) \langle R\rangle^5$ & $-100N^2\langle R\rangle^5$ \\
$6$ & $4N (14N^2+297N+1470) \langle R\rangle^6$ & $56N^3 \langle R\rangle^6$ \\
$7$ & $-4 N (287N^2 + 4046N + 20424) \langle R\rangle^7$ & $-1148 N^3 \langle
  R\rangle^7$ \\
$8$ & $N (431N^3 + 20594N^2 + 250576N + 1311648) \langle R\rangle^8$ &
  $431N^4 \langle R\rangle^8$ \\
$9$ & $-4 N (3453N^3 + 95021N^2 + 1089414N + 5957208) \langle R\rangle^9$ &
  $-13812N^4 \langle R\rangle^9$ \\
$10$ & $2 N (1971N^4 + 172657N^3 + 3737127N^2 + 42106610N + 241175496) \langle
  R\rangle^{10}$ & $3942N^5 \langle R\rangle^{10}$ \\
\end{tabular}
\end{ruledtabular}
\end{table*}

The density of eigenvalues $\tilde\lambda_n$ can be obtained from the resolvent
\cite{SVW01} $\tilde G(z) \equiv (z-\tilde A)^{-1}$. The density is given by the
spectral function
\be
\rho_\mathrm{all}(z) = -\frac{1}{\pi N}\, \mathrm{Im}\,\mathrm{Tr}\,\langle
\tilde G(z+i\eta)\rangle ,
\label{2.rhodef}
\ee
where $\eta\to0^+$ at the end of the calculation. The density
includes the exact zero eigenvalue so that we can write
\be
\rho_\mathrm{all}(z) = \frac{1}{N}\,\delta(z) + \frac{N-1}{N}\,\rho(z) ,
\ee
where $\rho(z)$ is the normalized density of nonzero eigenvalues.
In the limit of large $N$, the eigenvalue density $\rho_\mathrm{all}(z)\cong
\rho(z)$ only depends on the second moment $\langle \delta R^2\rangle$
of the distribution function $p$ of rates, at least as long as all moments of
$p$ exist. The proof is sketched in App.~\ref{app.a}. That the eigenvalue
distribution generically becomes independent of $p$ for large $N$ has been
conjectured by Mehta (conjecture 1.2.1 in Ref.~\cite{Meh04}). However, the
second
part of this conjecture, stating that the density of eigenvalues is the same as
for the Gaussian orthogonal ensemble (GOE), is not true for our ensemble.

Since the density of eigenvalues $\tilde\lambda_n$, $n>0$, of the shifted
matrices $\tilde A$ only depends on the second moment $\langle \delta
R^2\rangle$ for large $N$, we can obtain the large-$N$ behavior from any
distribution with that second moment. We choose the Gaussian distribution
\be
p_G(\tilde A_{ij}) = \frac{1}{\sqrt{2\pi \langle \delta
  R^2\rangle}}\, \exp\left(- \frac{\tilde A_{ij}^2}{2 \langle \delta
  R^2\rangle}\right) .
\ee
For this distribution together with the constraint $\sum_i \tilde
A_{ij}=0$, the eigenvalue density is known for large $N$ \cite{SMF03}: The
averaged resolvent is the solution of
\be
\langle\tilde G(z)\rangle = \frac{1}{\sqrt{N \langle \delta R^2\rangle}}\;
  g\!\left(\frac{z - N \langle \delta R^2\rangle\langle \tilde G(z)\rangle}
  {\sqrt{N \langle \delta R^2\rangle}} \right) ,
\label{2.largeN.1}
\ee
where
\be
g(z) \equiv \frac{1}{\sqrt{2\pi}} \int_{-\infty}^\infty \!\! dx\,
\frac{e^{-x^2/2}}{z-x} .
\label{2.largeN.2}
\ee
This integral can be evaluated,
\be
g(z) = \sqrt{\frac{\pi}{2}}\, z\,\sqrt{-\frac{1}{z^2}}\,
  e^{-z^2/2} \left(-2 + \mathrm{erfc}\, \frac{z^2\sqrt{-1/z^2}}{\sqrt{2}}
  \right) .
\ee
$g(z)$ has a cut along the whole real axis. The density $\rho(z)$ is thus
nonzero for all real $z$. Equations (\ref{2.rhodef}) and (\ref{2.largeN.1})
imply that $\sqrt{N \langle \delta R^2\rangle}\,\rho(z)$ is a universal
function of $z/\sqrt{N \langle \delta R^2\rangle}$.
The same distribution in the large-$N$ limit was found for adjacency
matrices \cite{BrR88,Fyo99}.
The corresponding result for the GOE is the well-known semicircle law
\cite{Wig57,Meh04}. It is worth pointing out that the different eigenvalue
density results only from the constraint $\sum_i \tilde
A_{ij}=0$.

We now study the eigenvalue density for the ESRE for finite $N$. We perform
Monte Carlo simulations by generating a number $n_r$ of realizations of matrices
from the ESRE for given $N$, shifted according to Eq.~(\ref{2.shiftA}). The
matrices are diagonalized and the eigenvalue with the numerically smallest
magnitude, which corresponds to $\tilde\lambda_0=0$, is dropped. The
eigenvalues are rescaled according to
$\tilde\lambda \to \tilde\lambda/\sqrt{N \langle \delta R^2\rangle}$. Finally,
histograms with 500 bins are generated.

\begin{figure}[htb]
\includegraphics[width=3.25in,clip]{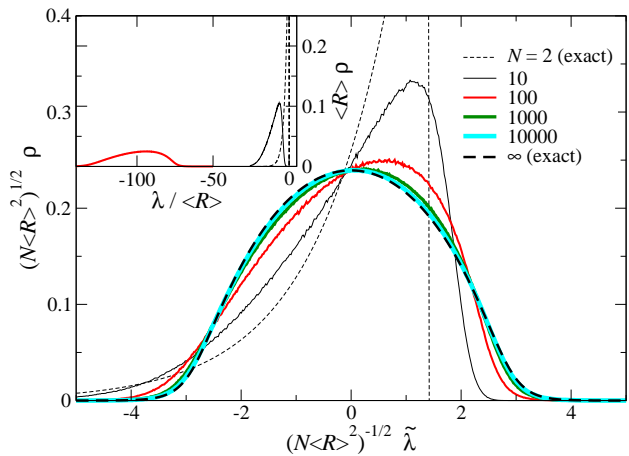}
\caption{\label{fig.SE.evaldensity}(Color online) Scaled density of nonzero
eigenvalues of shifted symmetric rate matrices $\tilde A=A-\langle
A\rangle$. The results for $N=2$ and $N\to\infty$ are exact, see text. The
curves for $N=10$, $100$, $1000$, $10000$ are
histograms with 500 bins for $10^7$ eigenvalues for matrices randomly chosen
from the ESRE. Inset: unscaled distribution of eigenvalues of the
unshifted matrices $A$ for $N=2$, $10$, $100$.}
\end{figure}

Results for $N=2$, $10$, $100$, $1000$, $10000$, and $\infty$ are shown in
Fig.~\ref{fig.SE.evaldensity}. For $N\to\infty$, we solve
Eq.~(\ref{2.largeN.1}). For $N=2$, the matrices have a single nonzero eigenvalue
$-2\tilde A_{12}$ with distribution following from
Eq.~(\ref{1.expdist}). For each of the other values of $N$, $n_rN=10^7$
eigenvalues have been generated. Figure \ref{fig.SE.evaldensity}
shows that the distribution changes smoothly from shifted exponential for $N=2$
to the known universal function for $N\to\infty$.
The inset in Fig.~\ref{fig.SE.evaldensity} shows the unscaled
eigenvalue density of the unshifted ESRE to illustrate that the mean scales
with $N$, whereas the width scales with $\sqrt{N}$.


While we have shown that nearly all nonzero eigenvalues lie in a narrow interval
around their mean for large $N$, the dynamics after a transient will be
dominated by the slowest process. The slowest non-stationary process is governed
by the eigenvalue $\lambda_1<0$ which is smallest in magnitude.
It is conceivable that matrices from the ESRE typically have an eigenvalue
$\lambda_1$ close
to zero. For example, $\lambda_1$ could scale with a lower power of $N$
compared to the mean $-N\langle R\rangle$. If the fraction of such anomalously
slow rates decreased for large $N$, they might not be visible in the
density plots in Fig.~\ref{fig.SE.evaldensity}.

\begin{figure}[htb]
\includegraphics[width=3.25in,clip]{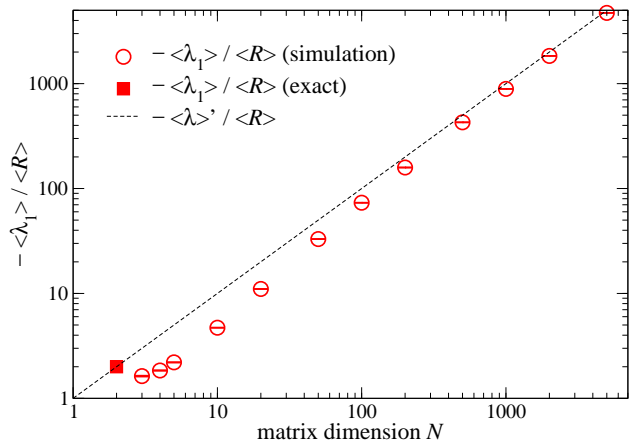}
\caption{\label{fig.SE.slow}(Color online) Average smallest in magnitude
eigenvalue, $\langle\lambda_1\rangle$, of matrices from the ESRE, as a function
of $N$. The open circles denote numerical results for $n_r=5000$ ($1000$)
realizations for $N\le 1000$ ($N\ge 2000$). Error bars denoting the statistical
errors are shown. The filled square denotes the
result $\lambda_1=-2\langle R\rangle$ for $N=2$. The dashed straight line
denotes the mean of nonzero eigenvalues, $-N\langle R\rangle$.}
\end{figure}

To check this, we plot the
mean $\langle\lambda_1\rangle$ as a function of
$N$ in Fig.~\ref{fig.SE.slow}. The average slowest rate
$|\langle\lambda_1\rangle|$ is significantly smaller than the average rate
$|\langle \lambda\rangle'|$ for small
$N$, as one would expect from the width $\sqrt{\mu_2}\propto\sqrt{N}$. On the
other hand, for large $N$, $|\langle\lambda_1\rangle|$ approaches
$|\langle \lambda\rangle'|$. Thus we do not find evidence for anomalously slow
processes. Instead, the slowest rate is consistent with the mean and width of
the eigenvalue distribution $\rho(\lambda)$.


\subsection{Eigenvalue correlations}

Since the eigenvalue density for the ESRE differs significantly
from the GOE, one might ask whether the correlations between eigenvalues
are also different. In the GOE, the distribution function of differences of
neighboring eigenvalues $\lambda$, $\lambda'$ approaches zero as
$|\lambda'-\lambda|$ for $\lambda'\to\lambda$.

\begin{figure}[htb]
\includegraphics[width=3.25in,clip]{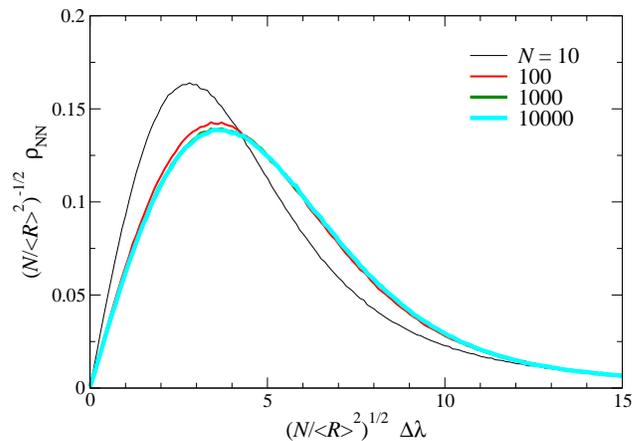}
\caption{\label{fig.SE.neighbors}(Color online) Scaled distribution of
nearest-neighbor separations $\Delta\lambda$ of nonzero eigenvalues for the ESRE
for $N=10$, $100$, $1000$, $10000$, from the same data sets as in
Fig.~\protect\ref{fig.SE.evaldensity}. The curve for $N=1000$ is nearly
obscured by the one for $N=10000$.}
\end{figure}

Figure \ref{fig.SE.neighbors} shows the distribution function
$\rho_\mathrm{NN}(\Delta\lambda)$ of separations
$\Delta\lambda\equiv \lambda_{n+1}-\lambda_n$ of neighboring eigenvalues
for the ESRE (here, the $\lambda_n$ are assumed to be ordered by value). The
zero eigenvalue $\lambda_0=0$ is excluded. Since the
width of the eigenvalue distribution scales as $\sqrt{N}$, while the number
of eigenvalues for a given realization scales as $N$, the typical separation
should scale as $1/\sqrt{N}$. We therefore rescale $\Delta\lambda \to
\sqrt{N/\langle R\rangle^2}\,\Delta\lambda$.
Figure \ref{fig.SE.neighbors} shows that the rescaled distribution approaches a
limiting form for $N\to\infty$. Furthermore, the distribution function
$\rho_\mathrm{NN}(\Delta\lambda)$ is linear in $\Delta\lambda$ for small
$\Delta\lambda$ for all $N$. Thus the distribution of nearest-neighbor
separations behaves essentially like for the GOE \cite{Meh04}. The constraint
(\ref{1.Aconstr}), which is
responsible for the deviation of the eigenvalue distribution from
the GOE result, does not have a comparably strong effect on the eigenvalue
correlations. The reason is very likely that the joint probability distribution
$\rho(\lambda_1,\lambda_2,\ldots,\lambda_{N-1})$ of the eigenvalues
\cite{Meh04},
while being complicated for the ESRE, does contain the factor
$\prod_{nn',0<n<n'} |\lambda_n-\lambda_{n'}|$, which determines the exponent
$\beta=1$ in $\rho_\mathrm{NN} \sim \Delta\lambda^\beta$.

\section{General rate-matrix ensemble}
\label{sec.EGRE}

We now turn to the ensemble of general, asymmetric rate matrices (EGRE).
Compared to the ESRE, it describes the opposite extreme of
independent rates $A_{ij}$ and $A_{ji}$ for forward and backward
transitions.

\subsection{Spectrum}
\label{sus.EGRE.spec}

As noted, there always exists an eigenvalue $\lambda_0=0$ with
left eigenvector $(1,1,\ldots,1)$. Other than for the symmetric case, the
corresponding right eigenvector is different. We are interested in the
distribution of the other
eigenvalues $\lambda_n$, $n=1,\ldots, N-1$, which are now complex with negative
real parts. We have already shown in Sec.~\ref{sec.ESRE} that the mean of
nonzero eigenvalues equals $-N\langle R\rangle$, see Eq.~(\ref{2.meanla}). We
shift the matrices according to Eq.~(\ref{2.shiftA}) so that they have zero
mean.

We define the expectation values
\be
\mu_m \equiv \langle \tilde\lambda^m \rangle'
  = \langle (\lambda -\langle\lambda\rangle)^m\rangle'
  = \langle (\lambda + N\langle R\rangle)^m\rangle'
\label{3.pseudomoms}
\ee
in analogy to the ESRE, but they are not the central moments of the
distribution of nonzero eigenvalues. Instead, the central moments have to be
defined for a two-dimensional distribution in the complex plane,
\be
\mu_{mn} \equiv \langle (\mathrm{Re}\,\lambda + N\langle R\rangle)^m\,
  (\mathrm{Im}\,\lambda)^n \rangle' .
\label{3.complexmoms}
\ee
Since the eigenvalues are real or form complex conjugate pairs, we have
$\mu_{mn}=0$ for odd $n$.
We show in App.~\ref{app.b} that the shifted eigenvalue distribution only
depends on the second moment $\langle \delta
R^2\rangle$ of $p$, like we found for the symmetric case.
We here call the $\mu_m$ in Eq.~(\ref{3.pseudomoms}) the
\emph{pseudomoments}. They are all
real, since the eigenvalues are real or form complex conjugate pairs.

The pseudomoments $\mu_m$ can be obtained in the same way as for symmetric
matrices. The results are different, since $\langle \tilde A_{ij} \tilde
A_{ji}\rangle = \langle\delta R^2\rangle$ for the symmetric case, whereas
$\langle \tilde A_{ij} \tilde A_{ji}\rangle = 0$ for the general case.
We present the pseudomoments $\mu_m$ up to $m=8$ for a general distribution
function $p(A_{ij})$ in Table \ref{tab.3} and up to $m=10$ for the
exponential distribution (EGRE) in Table \ref{tab.4}. The scaling of $\mu_m$
for even and odd $m$ and large $N$ is the same as for the ESRE. In the limit
$N\to\infty$, only the even pseudomoments survive. Interestingly, at least
up to $m=10$, these agree with the central moments of a real \emph{Gaussian}
distribution, $\mu_m^G = (m-1)!!\,(N\langle\delta R^2\rangle)^{m/2}$, where $n!!
= n(n-2)(n-4)\ldots$ is the double factorial. We show in App.~\ref{app.c} that
this identity holds for all even $m$.

\begin{table*}
\caption{\label{tab.3}Pseudomoments $\mu_m$, $m=2,\ldots,8$, of the
nonzero eigenvalues $\lambda$ for ensembles of general
rate matrices. The results hold
independently of the distribution function $p$ of rates $A_{ij}$, $i\neq j$,
as long as the moments exist.}
\begin{ruledtabular}
\begin{tabular}{cc}
$m$ & $\mu_m$ (general matrices, general distribution) \\
\hline
$2$ & $N \langle \delta R^2\rangle$ \\
$3$ & $-N \langle \delta R^3\rangle$ \\
$4$ & $N[3(N-1)\langle\delta R^2\rangle^2 + \langle\delta R^4\rangle]$ \\
$5$ & $-N[10(N-1)\langle\delta R^2\rangle\langle \delta R^3\rangle+
  \langle \delta R^5\rangle]$ \\
$6$ & $N[(15N^2-49N+38)\langle\delta R^2\rangle^3
  +10(N-1)\langle \delta R^3\rangle^2 + 15(N-1)\langle \delta R^2\rangle
  \langle\delta R^4\rangle + \langle\delta R^6\rangle]$ \\
$7$ & $-N[21(5N^2-17N+14) \langle \delta R^2\rangle^2 \langle\delta R^3\rangle
  +35(N-1) \langle\delta R^3\rangle\langle\delta R^4\rangle
  +21(N-1) \langle\delta R^2\rangle\langle\delta R^5\rangle
  +\langle\delta R^7\rangle]$ \\
$8$ & $N[3(35N^3-240N^2+551N-422)\langle\delta R^2\rangle^4
  +6(35N^2-121N+102) \langle\delta R^2\rangle^2 \langle\delta R^4\rangle
  +35(N-1) \langle\delta R^4\rangle^2$ \\
& $+56(N-1) \langle \delta R^3\rangle\langle\delta R^5\rangle
  +56(5N^2-18N+16) \langle\delta R^2\rangle\langle\delta R^3\rangle^2
  +28(N-1) \langle\delta R^2\rangle\langle\delta R^6\rangle
  +\langle\delta R^8\rangle]$ \\
\end{tabular}
\end{ruledtabular}
\end{table*}

\begin{table*}
\caption{\label{tab.4}Second column: pseudomoments $\mu_m$, $m=2,\ldots,10$,
of the nonzero eigenvalues $\lambda$ for ensembles of general rate
matrices, assuming an exponential distribution of rates (EGRE). Third column:
leading term of $\mu_m$ for large $N$.}
\begin{ruledtabular}
\begin{tabular}{ccc}
$m$ & $\mu_m$ (EGRE) & $\mu_m$ (EGRE, $N\gg 1$) \\
\hline
$2$ & $N \langle R\rangle^2$ & $N \langle R\rangle^2$ \\
$3$ & $-2N \langle R\rangle^3$ & $-2N \langle R\rangle^3$ \\
$4$ & $3N(N+2)\langle R\rangle^4$ & $3N^2 \langle R\rangle^4$ \\
$5$ & $-4N(5N+6) \langle R\rangle^5$ & $-20N^2\langle R\rangle^5$ \\
$6$ & $N (15N^2+126N+128) \langle R\rangle^6$ & $15N^3 \langle R\rangle^6$ \\
$7$ & $-6 N (35N^2 + 140N + 148) \langle R\rangle^7$ & $-210 N^3 \langle
  R\rangle^7$ \\
$8$ & $N (105N^3 + 2290N^2 + 6270N + 7476) \langle R\rangle^8$ &
  $105N^4 \langle R\rangle^8$ \\
$9$ & $-8 N (315N^3 + 2953N^2 + 6741N + 9018) \langle R\rangle^9$ &
  $-2520N^4 \langle R\rangle^9$ \\
$10$ & $N (945N^4 + 42494N^3 + 249174N^2 + 532840N + 774744) \langle
  R\rangle^{10}$ & $945N^5 \langle R\rangle^{10}$ \\
\end{tabular}
\end{ruledtabular}
\end{table*}

The eigenvalue distribution in the complex plane can be obtained from the
non-analyticities of the averaged resolvent $\langle\tilde G(z)\rangle =
\langle(z-\tilde A)^{-1}\rangle$ \cite{SVW01,Fei06}. However, unlike for
symmetric matrices, the non-analyticities are not limited to a branch cut along
the real axis.
%
%
For what follows, it is more convenient to employ the method of hermitization
\cite{Fei06}. We define the $2N\times 2N$ matrix
\be
\mathcal{H}(z,z^\ast) \equiv \left(\begin{array}{cc}
  0 & \tilde A - zI \\
  \tilde A^T - z^\ast I & 0
  \end{array}\right) ,
\label{3.Hdef}
\ee
where $\tilde A^T$ is the transpose of $\tilde A$. $\mathcal{H}(z,z^\ast)$ is
hermitian for any complex $z$. With the resolvent of $\mathcal{H}$,
\be
\mathcal{G}(\eta;z,z^\ast) \equiv \frac{1}{\eta-\mathcal{H}(z,z^\ast)} ,
\ee
the density of eigenvalues in the complex plane is \cite{Fei06}
\be
\rho_\mathrm{all}(x,y) = \frac{1}{\pi N}\, \frac{\partial}{\partial z^\ast}\,
  \mathrm{Tr}_{2N}
  \left(\begin{array}{cc}
    0 & I \\ 0 & 0
  \end{array}\right)
  \langle\mathcal{G}(0;z,z^\ast)\rangle ,
\label{3.rhofullhermsd}
\ee
where $z=x+iy$, the derivative with respect to $z^\ast$ is to be taken
with $z$ fixed, and $\mathrm{Tr}_{2N}$ denotes the trace over a
$2N\times 2N$ matrix. Using this representation, we show that for large
$N$ the eigenvalue density only depends on the second central moment
$\langle\delta R^2\rangle$ of the distribution of rates $A_{ij}$. The proof is
sketched in App.~\ref{app.b}. Edelman \textit{et al.}\ \cite{EKS94} have
conjectured that this is generically the case for asymmetric matrices.



We now present numerical results for $\rho(x,y)$ for the EGRE, as a function of
the matrix dimensions $N$. As above, $\rho_\mathrm{all}$
contains all eigenvalues, whereas $\rho$ excludes the exact zero. We will
compare the results to the Ginibre
ensemble of real asymmetric matrices with Gaussian distribution of components
(Ginibre orthogonal ensemble, GinOE)
\cite{Gin65,Gir84,EKS94,Bai97,Ede97,KaA05,SoW08}, which is
the closest relative of the EGRE that has been studied in detail.

\begin{figure}[htb]
\includegraphics[width=2.70in,clip]{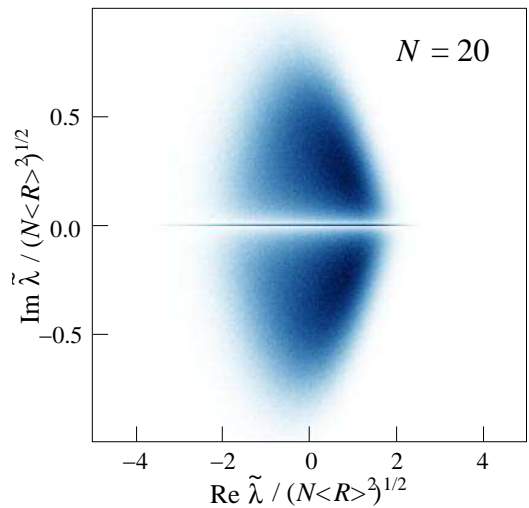}
\caption{\label{fig.GE.alleval20}(Color online) Scaled distribution function
of nonzero eigenvalues $\tilde\lambda$ of shifted general rate matrices
$\tilde A$ of dimension $N=20$. More specifically, a
two-dimensional histogram with $500\times500$ bins was populated for $n_r$
matrices randomly chosen from the EGRE, with $n_rN=4\times 10^7$.}
\end{figure}

As observed above, the eigenvalues $\tilde\lambda$ of $\tilde A$ can be either
real or form complex conjugate pairs. The numerical simulations show that
both types of eigenvalues indeed occur. A typical eigenvalue density is shown
in Fig.~\ref{fig.GE.alleval20} for $N=20$. We assume that the square
root of the second pseudomoment, $\sqrt{\mu_2}=\sqrt{N \langle R\rangle^2}$,
describes the typical
width of the distribution and rescale the eigenvalue density accordingly.
The real and complex eigenvalues are clearly visible. Here and in the following
``complex'' should be understood as ``not real.''
Figure \ref{fig.GE.alleval20} already suggests that the distribution of nonzero
eigenvalues of $A$ becomes a narrow peak around $-N\langle R\rangle$ for large
$N$, like for the ESRE. We return to this point below.

The question arises of what fraction $f_\mathbb{R}$ of the nonzero eigenvalues
are real. For the GinOE, this fraction is known analytically
\cite{EKS94}. (The probability of finding exactly
$N_\mathbb{R}$ real eigenvalues for $N\times N$ matrices from the GinOE is also
known \cite{KaA05}.) Edelman
\textit{et al.}\ \cite{EKS94} derive various equivalent expressions for the
expected number of real eigenvalues, $\langle N_\mathbb{R}\rangle$, from which
we obtain $f_\mathbb{R}^\mathrm{GinOE} = \langle N_\mathbb{R}\rangle/N$.
We here quote an expression in terms of the hypergeometric
function ${}_2F_1$ \cite{EKS94}:
\be
f_\mathbb{R}^\mathrm{GinOE} = \frac{1}{2N} + \sqrt{\frac{2}{\pi}}\,
  \frac{\Gamma(N+1/2)}{\Gamma(N+1)}\, {}_2F_1\left(1,-\frac12;N;\frac12\right) .
\label{3.fhyper}
\ee
For large $N$, this becomes \cite{EKS94}
\be
f_\mathbb{R}^\mathrm{GinOE} \cong \sqrt{\frac{2}{\pi N}} .
\ee
For the GinOE, the fraction of real eigenvalues thus asymptotically decays with
a simple exponent of $-1/2$.

\begin{figure}[htb]
\includegraphics[width=3.25in,clip]{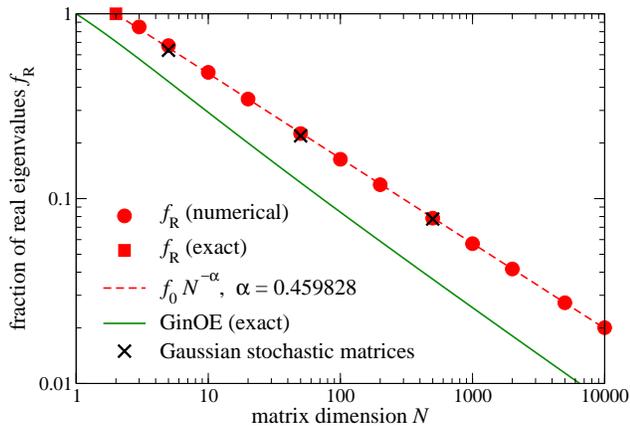}
\caption{\label{fig.GE.realfrac}(Color online) Fraction $f_\mathbb{R}$ of
nonzero eigenvalues that are real, as a function of $N$ for the EGRE. The solid
circles denote numerical values obtained for $n_r$ realizations with
$n_rN=4\times 10^7$ for $N\le 2000$, $n_rN=10^7$ for $N=5000$, and $n_rN=4\times
10^5$ for $N=10000$. The solid square represents the exact result
$f_\mathbb{R}=1$
for $N=2$. The dashed line denotes a power law $f_0 N^{-\alpha}$ fitted to the
two points for $N=2000$ and $N=5000$. The solid line is the exact result for the
GinOE, Eq.~(\protect\ref{3.fhyper}). The crosses denote numerical results for
ensembles of rate matrices with Gaussian instead of exponential
distribution of rates $A_{ij}$, $i\neq j$.}
\end{figure}

Figure \ref{fig.GE.realfrac} shows the fraction $f_\mathbb{R}$ as a function
of $N$ for the EGRE and for comparison the exact result for the GinOE. For
$N=2$, $f_\mathbb{R}$ must be unity, since the single nonzero eigenvalue cannot
be a complex conjugate pair. The results clearly
differ from the GinOE and decay more slowly for large $N$. A fit of a
power law $f_\mathbb{R} \sim f_0 N^{-\alpha}$ to the data points for $N=2000$
and $5000$ is also included in Fig.~\ref{fig.GE.realfrac}. We obtain
$f_0\approx 1.37$ and $\alpha\approx 0.460$. The large-$N$ behavior is
inconsistent with the exponent $1/2$ found for the GinOE. This is remarkable,
since all other scaling relations we have so far found, as well as the ones for
the GinOE, only contain integer powers of $\sqrt{N}$. Physically, this means
that the fraction of eigenvectors describing purely exponentially
decaying deviations from the stationary state scales with a nontrivial power
$-\alpha$ of the number of states.

To pinpoint the origin of the anomalous scaling, we have also
evaluated $f_\mathbb{R}$ for ensembles of matrices of dimension
$N=5$, $50$, $500$ satisfying the constraint (\ref{1.Aconstr}), but with
\emph{Gaussian} distribution of rates $A_{ij}$, $i\neq j$. This is the
asymmetric analogue
of the symmetric ensemble studied by St\"aring \textit{et al.}\ \cite{SMF03}.
The results are shown as crosses in Fig.~\ref{fig.GE.realfrac}. They clearly
approach the EGRE results for large $N$, not the GinOE. It is thus the
constraint (\ref{1.Aconstr}) that leads to the anomalous scaling.

\begin{figure}[htb]
\includegraphics[width=3.25in,clip]{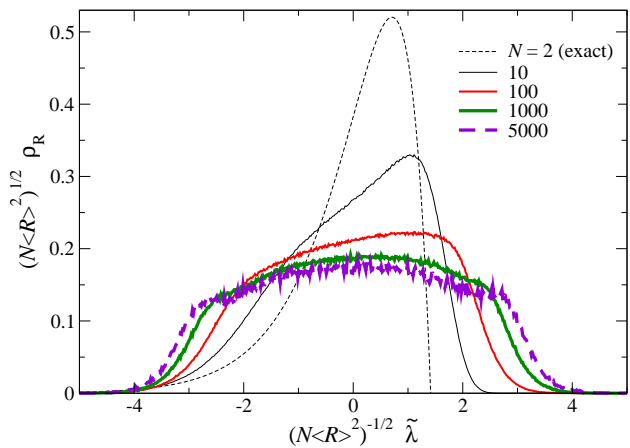}
\caption{\label{fig.GE.realeval}(Color online) Scaled density of nonzero real
eigenvalues of shifted general rate matrices $\tilde A$.
The curve for $N=2$ is exact. The curves for $N=10$, $100$, $1000$, $5000$ are
histograms with 500 bins for a total number $n_rN=4\times 10^7$ ($10^7$) of
eigenvalues (including complex ones) for $N\le 1000$ ($5000$) for matrices
randomly chosen from the EGRE.}
\end{figure}

In the following, we will consider the real and complex eigenvalues separately.
Figure \ref{fig.GE.realeval} shows the density $\rho_\mathbb{R}$ of shifted real
nonzero eigenvalues $\tilde\lambda$, normalized to unity and rescaled with the
square root of the pseudomoment $\sqrt{\mu_2}=\sqrt{N\langle R\rangle^2}$,
for $N=2$, $10$, $100$, $1000$, $5000$. For $N=2$, the single nonzero eigenvalue
is $\tilde\lambda=-\tilde A_{12}-\tilde A_{21}$. In the
EGRE, its distribution function is
$\rho_\mathbb{R}(\tilde\lambda) = (2/\langle R\rangle - \tilde\lambda/\langle
R\rangle^2)\,\exp(\tilde\lambda/\langle R\rangle-2)$ for $\tilde\lambda\le
2\langle R\rangle$ and zero otherwise. For the other values of $N$,
Fig.~\ref{fig.GE.realeval} shows numerical results. The noise increases for
large $N$, not only because $n_rN$ was smaller for
$N=5000$ but also because $f_\mathbb{R}$ decreases with
increasing $N$. It is obvious however that the distribution for large $N$ is
quite different from the eigenvalue density for the ESRE,
Fig.~\ref{fig.SE.evaldensity}.

The distribution clearly becomes more symmetric for $N\to\infty$, as it must,
since the large-$N$ result only depends on the width of the distribution of
rates $A_{ij}$. There is an indication that the distribution develops
non-analyticities with sudden changes of slope in the limit
$N\to\infty$. This is not unexpected, since the scaled distribution
of real eigenvalues of the GinOE is uniform on the interval $[-1,1]$ and zero
otherwise \cite{EKS94,SoW08} and thus also shows non-analyticities. Compared to
the ESRE (Fig.~\ref{fig.SE.evaldensity}), the convergence to the large-$N$ limit
is slower for the EGRE (Fig.~\ref{fig.GE.realeval}). In fact,
from Fig.~\ref{fig.GE.realeval} we cannot exclude the possibility
that the width scales with an anomalous power of $N$, different from $1/2$.



\begin{figure}[htb]
\includegraphics[width=2.70in,clip]{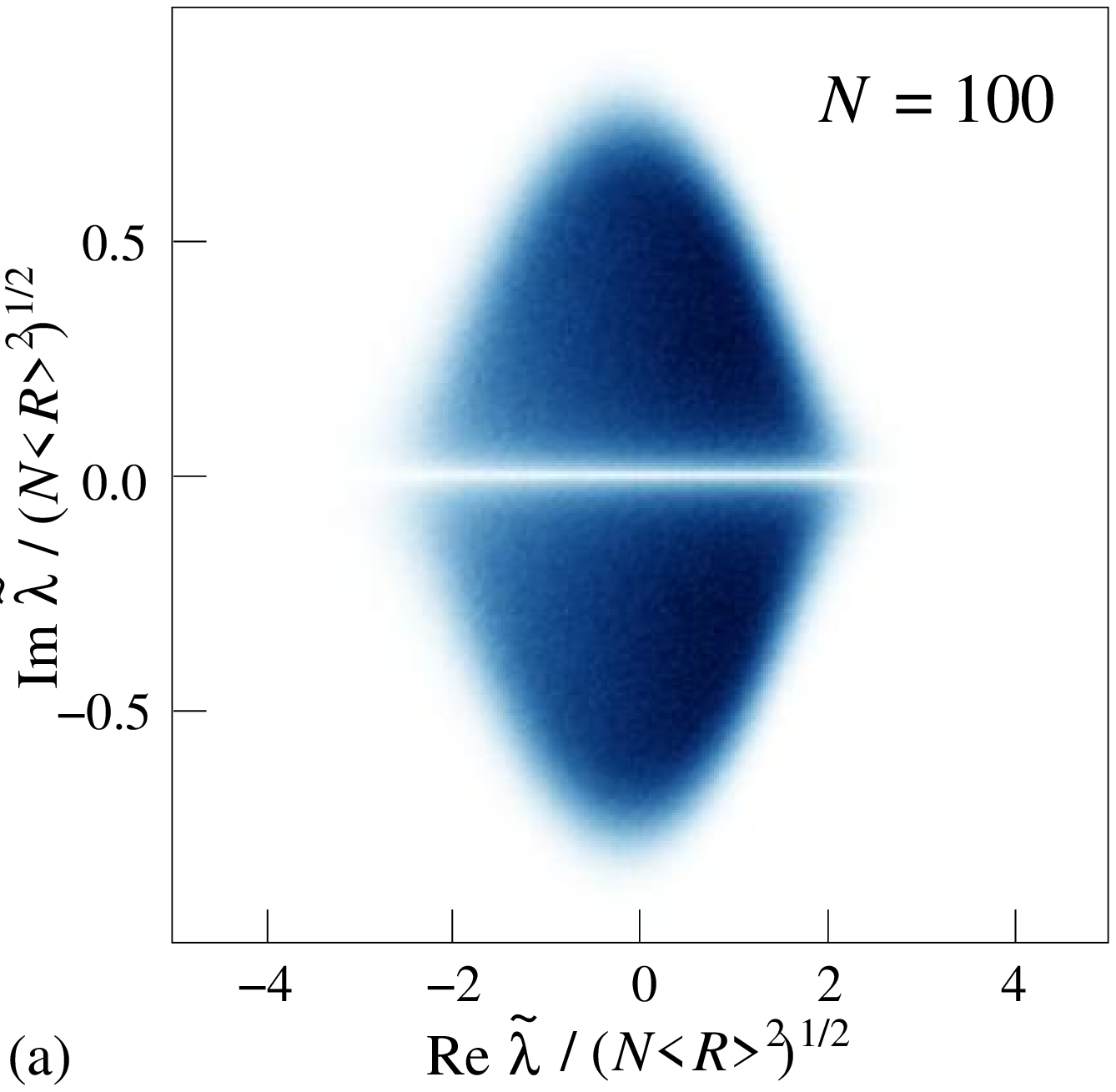}\\[2ex]
\includegraphics[width=2.70in,clip]{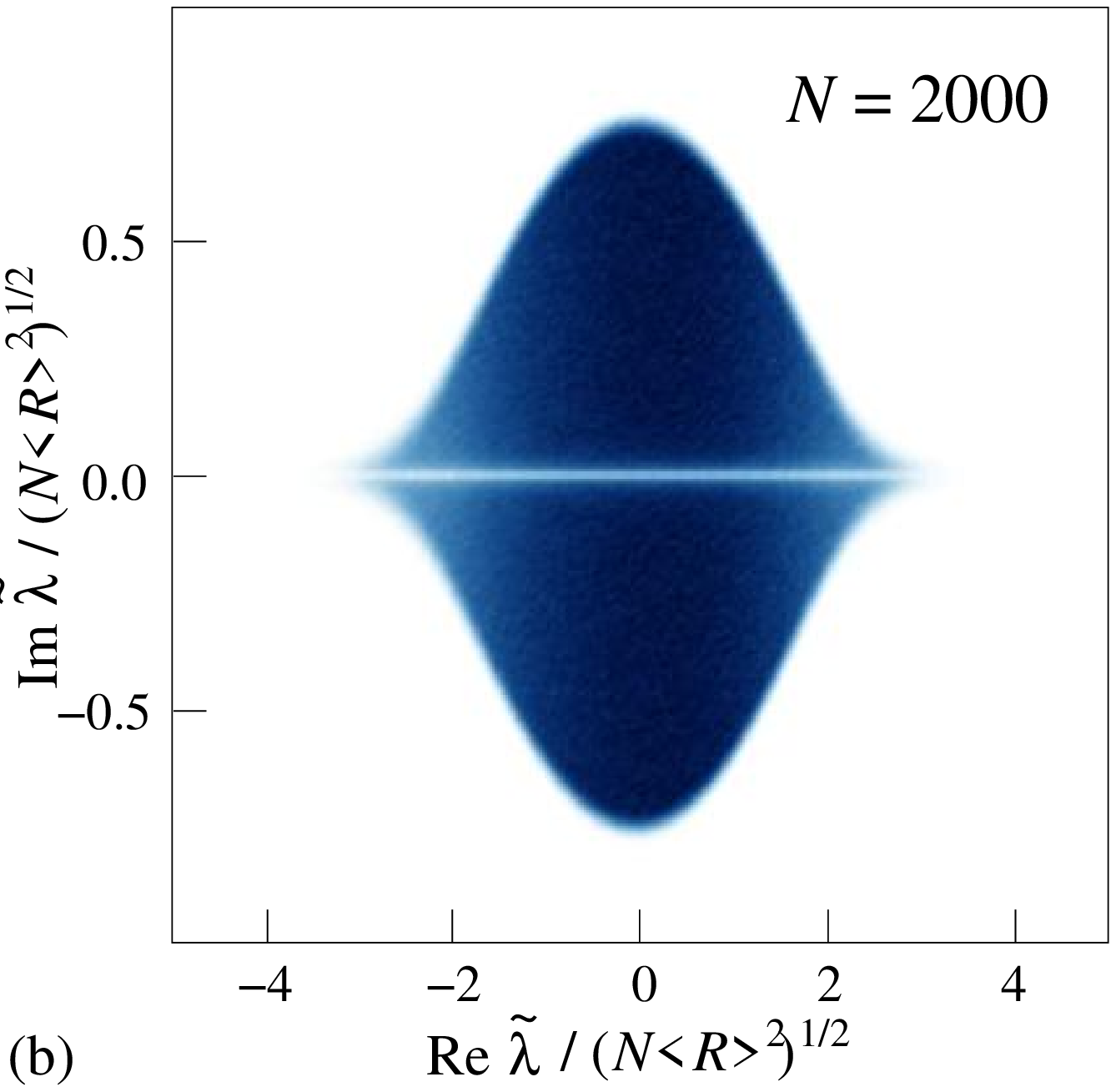}
\caption{\label{fig.GE.compeval}(Color online) Scaled distribution function of
complex eigenvalues
$\tilde\lambda$ of shifted general rate matrices $\tilde A$
of dimension (a) $N=100$ and (b) $N=2000$. Specifically,
two-dimensional histograms with $500\times500$ bins were populated for $n_r$
matrices randomly chosen from the EGRE, where $n_rN=4\times
10^7$. Note the different scales of the axes.}
\end{figure}

Turning to complex eigenvalues, we note that for large $N$ nearly all
eigenvalues belong to this class, since the fraction $f_\mathbb{R}$ of real
eigenvalues approaches zero. We plot their distribution function
$\rho_\mathbb{C}$ in the complex plane for $N=100$ and $2000$ in
Fig.~\ref{fig.GE.compeval}. The scaled distribution for $N=5000$ is virtually
indistinguishable from the one for $N=2000$. From
Figs.~\ref{fig.GE.alleval20} and \ref{fig.GE.compeval}, we see that the
distribution becomes more symmetric with respect to inversion of the real part
as $N$ increases.

The widths of the distribution in the real
direction, $\sqrt{\mu_{2,0}}$, and in the imaginary direction,
$\sqrt{\mu_{0,2}}$, see Eq.~(\ref{3.complexmoms}), both scale with
$\sqrt{N\langle R\rangle^2}$. This means that the typical decay rate is
$\langle\lambda\rangle'=N\langle R\rangle$, whereas the typical oscillation
frequency is of the order of $\sqrt{N}\langle R\rangle$. For large $N$ it will
thus be difficult to observe the oscillations.

\begin{figure}[htb]
\includegraphics[width=3.25in,clip]{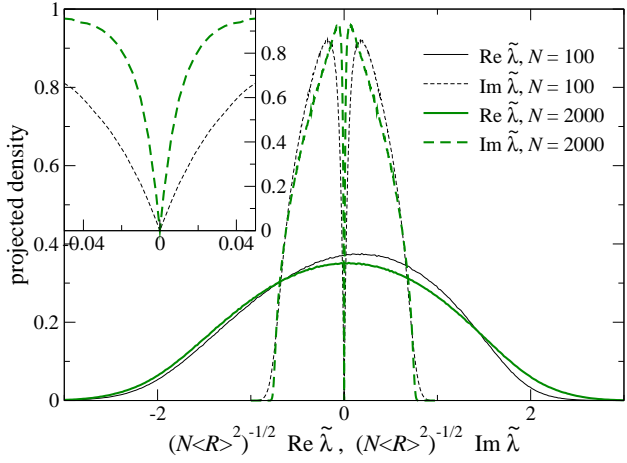}
\caption{\label{fig.GE.projcompeval}(Color online) Scaled density of the real
part (solid lines) and the imaginary part (dashed lines) of complex eigenvalues
of shifted general rate matrices with $N=100$ and $N=2000$. The curves are
projections of the data shown in Fig.~\protect\ref{fig.GE.compeval} onto the
real and imaginary axes. The inset shows the scaled density of the
imaginary part around zero.}
\end{figure}

It is instructive to compare the distribution to the one for the GinOE. For
the GinOE, the distribution function $\rho_\mathbb{C}$ of complex eigenvalues
for finite $N$ has been obtained by Edelman \cite{Ede97} in terms of a finite
sum of $N-1$ terms, which can be rewritten as a simple integral \cite{SoW08}.
The distribution function $\rho_\mathbb{C}$ is found to contain a factor
$|\mathrm{Im}\,\tilde\lambda|$, showing that the density goes to
zero linearly for $\tilde\lambda$ approaching the real axis. Complex
eigenvalues
are thus repelled by the real axis with a characteristic exponents of unity.
Figures \ref{fig.GE.alleval20} and \ref{fig.GE.compeval} clearly show that
complex eigenvalues are also repelled by the real axis for the EGRE. In
Fig.~\ref{fig.GE.projcompeval} we plot the density of complex eigenvalues,
projected onto the real and imaginary axes, for $N=100$ and $N=2000$. We
observe that for the EGRE the complex eigenvalues are repelled by the real axis
with the same exponent of unity. We note that the distribution of the real part
of complex eigenvalues is distinct from
both the distribution of real eigenvalues, Fig.~\ref{fig.GE.realeval}, and the
distribution of eigenvalues for the ESRE, Fig.~\ref{fig.SE.evaldensity}.

For the GinOE, the scaled distribution approaches a uniform
distribution on the unit disk in the complex plane for $N\to\infty$. This
was conjectured by Girko \cite{Gir84} for an arbitrary distribution of
components with zero mean and proven by Bai \cite{Bai97}. The EGRE result
is clearly much more complicated. The histograms for various values of
$N$ suggest that the distribution function $\rho_\mathbb{C}$ does not become
uniform in a bounded region for $N\to\infty$, although it does appear to
develop non-analyticities, which show up as high-contrast edges in
Fig.~\ref{fig.GE.compeval}(b).

We now return to the moments of the distribution function
$\rho$ of \emph{all} nonzero eigenvalues of $\tilde A$. The moments
$\mu_{mn}$, Eq.~(\ref{3.complexmoms}), and the pseudomoments $\mu_m$,
Eq.~(\ref{3.pseudomoms}), are related. This is easily seen for $\mu_2$:
\be
\mu_2 = \langle \tilde\lambda^2 \rangle'
  = \langle (\mathrm{Re}\,\tilde\lambda)^2 + 2i\,\mathrm{Re}\,\tilde\lambda\,
  \mathrm{Im}\,\tilde\lambda - (\mathrm{Im}\,\tilde\lambda)^2 \rangle .
\ee
Since the second term vanishes, we obtain $\mu_2 = \mu_{2,0} - \mu_{0,2}$.
Now $\mu_{2,0}$ contains contributions from the real and the complex
eigenvalues, while $\mu_{0,2}$ only depends on the complex eigenvalues. We can
write
\be
\mu_2 = f_\mathbb{R}\, \mu_2^\mathbb{R} + (1-f_\mathbb{R})\,
\mu_{2,0}^\mathbb{C}
  - (1-f_\mathbb{R})\, \mu_{0,2}^\mathbb{C} ,
\ee
where the superscript $\mathbb{R}$ or $\mathbb{C}$ refers to the moments of the
distributions of real and complex eigenvalues, respectively. In the limit of
large $N$ we know that $f_\mathbb{R}\to 0$ and $\mu_2 \cong N\langle
R\rangle^2$. This means that the scaled distribution in the complex plane must
be anisotropic: The width in the imaginary direction must be smaller by a value
of the order of unity than in the real direction, unlike for the GinOE.
This is seen in Fig.~\ref{fig.GE.compeval}.

For arbitrary even $m$, the relation reads
\ba
\mu_m & = & \!\!\! \sum_{n=0\,\mathrm{even}}^m (-1)^{n/2}
  \left({m\atop n}\right)
  \mu_{m-n,n} \nonumber \\
& = & f_\mathbb{R}\, \mu_m^\mathbb{R}
  + (1-f_\mathbb{R}) \!\! \sum_{n=0\,\mathrm{even}}^m \! (-1)^{n/2}
  \left({m\atop n}\right)
  \mu_{m-n,n}^\mathbb{C} . \nonumber \\
& &
\label{3.musumrule}
\ea
We recall that the $\mu_m$ for small $m$ are known for all $N$, see Table
\ref{tab.4}. For large $N$, we have the asymptotically exact expression
(\ref{app.c.pseudomres}), which can be written as
$\mu_m \cong (m-1)!!\, N^{m/2}\, \langle R\rangle^m$. Hence,
we find asymptotically exact sum rules for all even orders $m$.

\begin{figure}[htb]
\includegraphics[width=3.25in,clip]{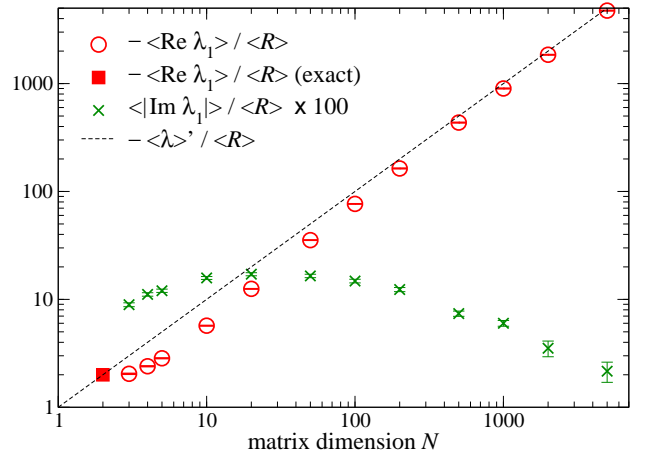}
\caption{\label{fig.GE.slow}(Color online) Typical real and imaginary parts of
the eigenvalue belonging to the slowest non-stationary process, as functions of
$N$. The open circles denote the average smallest in magnitude
real part of eigenvalues, $\langle\mathrm{Re}\,\lambda_1\rangle$, of matrices
from the EGRE. The data are numerical
results for $n_r=5000$ ($1000$) realizations for $N\le
1000$ ($N\ge 2000$). The crosses denote the average magnitude of the imaginary
part of the same eigenvalues, $\langle|\mathrm{Im}\,\lambda_1|\rangle$, scaled
$\times 100$. Error bars denoting the statistical errors are shown. The
filled square denotes the exact
result $\lambda_1=-2\langle R\rangle$ for $N=2$. The dashed line
denotes the mean of nonzero eigenvalues, $-N\langle R\rangle$.}
\end{figure}

To end this section, we again consider the slowest process. The dynamics at late
times is typically governed by
the eigenvalue $\lambda_1$ with the largest (smallest in magnitude) real part.
In Fig.~\ref{fig.GE.slow} we show the mean of the real part
$\mathrm{Re}\,\lambda_1$ and of the magnitude of the imaginary part,
$|\mathrm{Im}\,\lambda_1|$ for random matrices from the EGRE, as functions of
$N$. The behavior of the real part, i.e., the rate, is very similar to the ESRE.
Again, the slowest rate is consistent with the mean and width of
the eigenvalue distribution $\rho(\lambda)$. The typical imaginary part of
$\lambda_1$, i.e., the oscillation frequency, decreases for large $N$,
mainly because the probability of $\lambda_1$ being real
increases. While the fraction of real eigenvalues approaches
zero for large $N$, the eigenvalue with the largest real part becomes more
likely to be real.

\subsection{Eigenvalue correlations}

The eigenvalue density for the EGRE is quite different from the GinOE. Like for
the ESRE, we again ask whether the eigenvalue correlations are also different.
We consider the real and complex eigenvalues separately. The main effect of
correlations between real and complex eigenvalues is seen in
Fig.~\ref{fig.GE.projcompeval}: The complex eigenvalues are repelled by the real
axis with a characteristic exponent of unity.

\begin{figure}[htb]
\includegraphics[width=3.25in,clip]{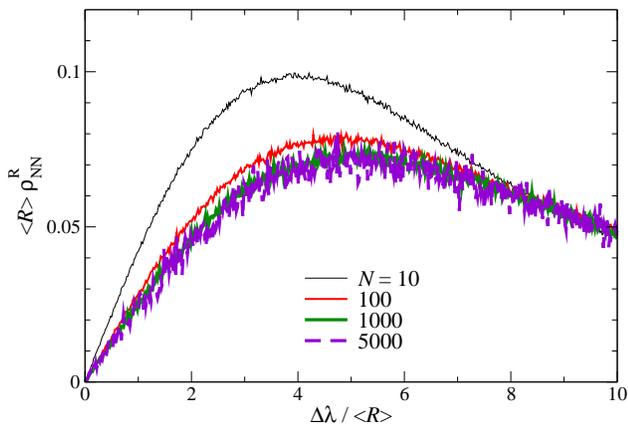}
\caption{\label{fig.GE.realneighbors}(Color online) Distribution of
nearest-neighbor separations $\Delta\lambda$ of nonzero real eigenvalues for the
EGRE for $N=10$, $100$, $1000$, $5000$ for the same data sets as in
Fig.~\protect\ref{fig.GE.realeval}. The axes are not rescaled with
a power of $N$.}
\end{figure}

Figure \ref{fig.GE.realneighbors} shows the distribution function
$\rho_\mathrm{NN}^\mathbb{R}(\Delta\lambda)$ of separations of neighboring real
eigenvalues. Note that the distribution is not rescaled with a power of $N$.
The typical separation of real eigenvalues depends only weakly on
$N$ for large $N$ for the EGRE, whereas it scales with $N^{-1/2}$ for
the ESRE. This can be understood as follows: The expected number of real
eigenvalues of a randomly chosen matrix is $Nf_\mathbb{R} \sim N^{1-\alpha}$,
while the width of their distribution scales with $N^{1/2}\langle R\rangle$.
Consequently, the typical nearest-neighbor separation should scale with
$N^{\alpha-1/2}\langle R\rangle$. Since
$\alpha$ is close to $1/2$, we obtain a weak dependence on $N$.
The dependence on separation $\Delta\lambda$ is again linear for small
$\Delta\lambda$, though. Thus real eigenvalues repel each other with a
characteristic exponent of unity, like for the GinOE \cite{SoW08}.

\begin{figure}[htb]
\includegraphics[width=2.70in,clip]{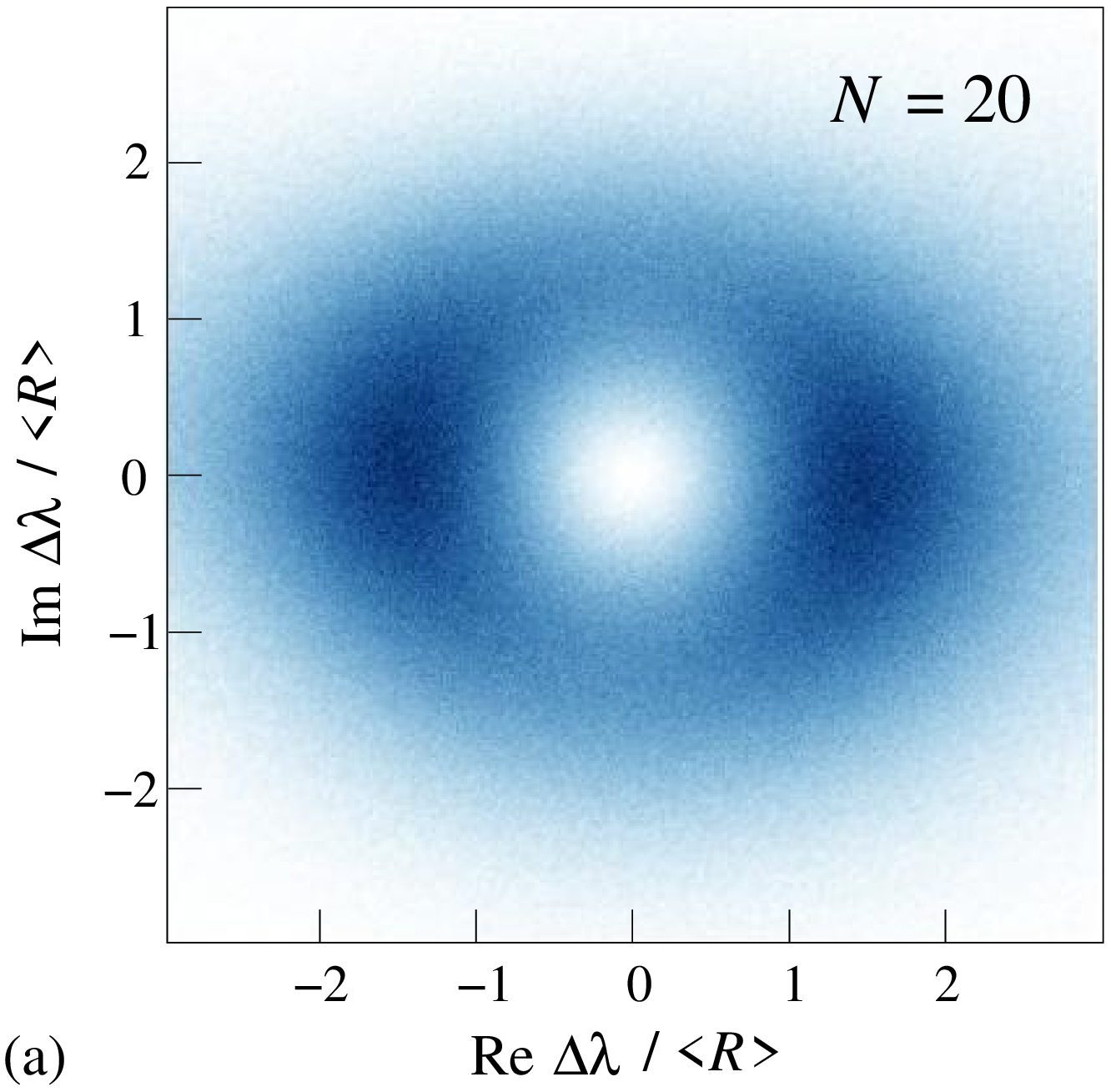}\\[2ex]
\includegraphics[width=2.70in,clip]{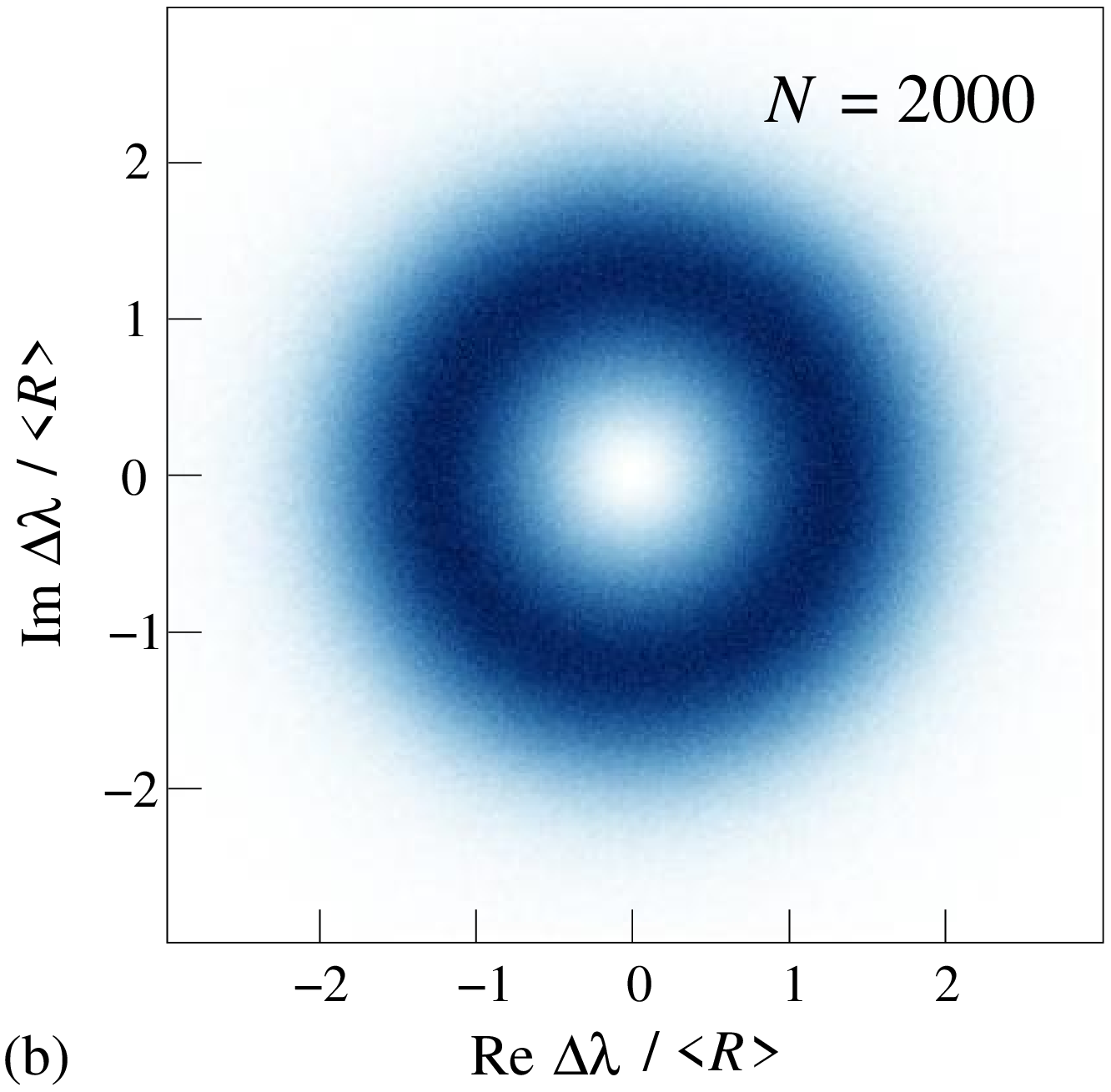}
\caption{\label{fig.GE.compneighbors}(Color online) Distribution
function of complex differences $\Delta\lambda$ of neighboring eigenvalues with
positive imaginary part for the EGRE for (a) $N=20$ and
(b) $N=2000$.}
\end{figure}

In Figs.~\ref{fig.GE.compneighbors}(a) and (b), we plot the distribution
function $\rho_\mathrm{NN}^\mathbb{C}(\Delta\lambda)$ of complex differences of
neighboring complex eigenvalues with positive imaginary part for $N=20$
and $N=2000$. More specifically, for each eigenvalue
$\tilde\lambda$ with positive imaginary part, we determine the eigenvalue
$\tilde\lambda'$ with positive imaginary part that minimizes
$|\tilde\lambda'-\tilde\lambda|$. We then collect the
complex differences $\Delta\lambda\equiv \tilde\lambda'-\tilde\lambda$
of all such pairs in a two-dimensional histogram.
The eigenvalues with negative imaginary part just form a mirror
image. Correlations between eigenvalues
with positive and negative imaginary parts are dominated by their repulsion by
the real axis and a $\delta$-function
from complex conjugate pairs and are not considered further.

Since the fraction of complex eigenvalues approaches unity for $N\to\infty$, the
number of complex eigenvalues of a chosen matrix scales with $N$. The widths of
the distribution in both the real and the imaginary direction scale with
$\sqrt{N}$, see Fig.~\ref{fig.GE.compeval}. The typical nearest-neighbor
distance should thus approach a constant for large $N$. This is indeed seen in
Fig.~\ref{fig.GE.compneighbors}.

\begin{figure}[htb]
\includegraphics[width=3.25in,clip]{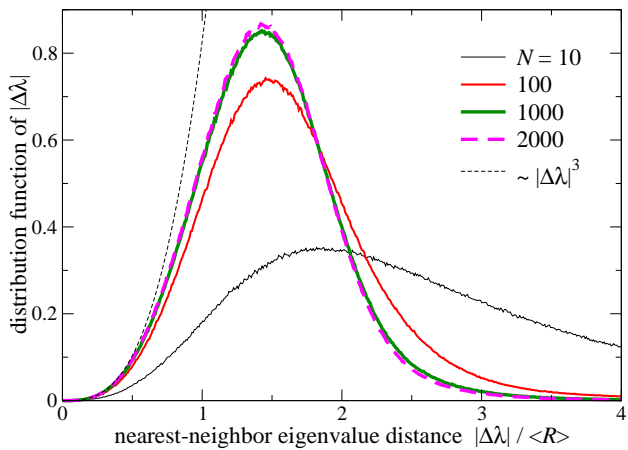}
\caption{\label{fig.GE.magneighbors}(Color online) Distribution function of
nearest-neighbor distances $|\Delta\lambda|$ for various values of $N$. The
dashed curve shows a power law $\propto |\Delta\lambda|^3$.}
\end{figure}

We observe that the distribution of differences becomes rotationally symmetric
for large $N$. This is perhaps surprising since the
distribution of the eigenvalues themselves is far from symmetric, see
Fig.~\ref{fig.GE.compeval}. Also, small differences are suppressed, i.e., the
eigenvalues repel each other. To find the characteristic exponent, we
plot the distribution of the magnitudes
$|\Delta\lambda| = |\tilde\lambda'-\tilde\lambda|$ of differences
of neighboring eigenvalues in Fig.~\ref{fig.GE.magneighbors}. We
observe that the distribution behaves like $|\Delta\lambda|^3$ for small
$|\Delta\lambda|$. Together with the rotational symmetry this implies that the
two-dimensional distribution in the complex plane,
Fig.~\ref{fig.GE.compneighbors}(b), approaches zero like
$|\Delta\lambda|^2$. The exponent of two is the same as for the GinOE
\cite{SoW08}. We conclude that the constraint (\ref{1.Aconstr}) and the
exponential distribution of rates in the EGRE do not change the repulsion of
neighboring eigenvalues compared to the GinOE, while the eigenvalue density is
very different. The origin of this is likely the same as to the ESRE: The
correlations are governed by ``local'' properties of the joint distribution
function of eigenvalues, which are not strongly affected by the constraint.

\section{Conclusions}
\label{sec.conc}

We have applied RMT to the transition-rate matrix $A$, i.e., the
matrix of coefficients in the Pauli master equation (\ref{1.stocheq}). This
allows us to obtain statistical properties of the spectrum, in analogy to RMT
for Hamiltonians. For the master equation, the eigenvalues describe the
decay, and, in the case of complex eigenvalues, the superimposed oscillations,
of probability eigenvectors.


The resulting random-matrix ensembles are different from the standard ensembles
for Hamiltonians, since $A$ is real but in general not symmetric and since the
conservation of probability imposes the constraint $\sum_i A_{ij}=0$ for all
$j$, Eq.~(\ref{1.Aconstr}). Although this constraint represents only
$N$ conditions for of the order of $N^2$ matrix components, its
consequences persist for large $N$.


A further difference to the standard ensembles is that the off-diagonal
components of the rate matrix represent rates and thus must be
non-negative. We have assumed an exponential distribution. The results in the
large-$N$ limit are found to be independent of the distribution of rates,
though.

We have considered both symmetric and general, asymmetric rate matrices.
The first case corresponds to systems where the rates for transitions from any
state $i$ to any other state $j$ and from $j$ to $i$ are identical. In the
second case, these rates are assumed to be independent. In both
cases, all nonzero eigenvalues form a narrow distribution of width
proportional to $\sqrt{N}$ around their mean, $-N\langle R\rangle$, where
$\langle R\rangle$ is the average transition rate. Thus for not too small $N$,
nearly all deviations from the stationary state decay on the same time scale
$1/N\langle R\rangle$. For both cases, we have found that the slowest
non-stationary state, which dominates the dynamics at late times,
typically also decays on the same time scale. We have derived exact expressions
for the expectation values of $m$-th powers of the nonzero eigenvalues, for
small $m$, for both cases.



For \emph{symmetric} rate
matrices, the density of eigenvalues has been studied numerically as a function
of $N$ and found to approach the same limiting form for $N\to\infty$ as obtained
earlier for Gaussian and two-valued distributions
\cite{BrR88,Fyo99,SMF03}, but very different from the semi-circle law for the
GOE \cite{Wig57,Meh04}. This difference is due to the constraint
(\ref{1.Aconstr}). On the other hand, the correlations between eigenvalues are
dominated by a repulsion with a characteristic exponent of
unity, as for the GOE.


For \emph{general} rate matrices,
we have numerically studied the eigenvalue density in the complex plane as a
function of $N$. For large $N$, it approaches a non-trivial distribution
different from the disk found for the GinOE \cite{Gir84,Bai97}. Interestingly,
the fraction of nonzero eigenvalues that are real decays as $N^{-\alpha}$
with an anomalous exponent $\alpha\approx 0.460$, unlike
for the GinOE, where $\alpha=1/2$. Thus the
fraction of eigenvectors describing purely exponentially decaying
deviations from the stationary state scales with a nontrivial power
of the number of possible states. Both the non-trivial distribution and the
anomalous scaling for large $N$ are due to the constraint
(\ref{1.Aconstr}). The density of real eigenvalues is also different from the
GinOE. We have obtained simple analytical results for the expectation values
$\langle(\lambda-\langle\lambda\rangle')^m\rangle'
= (m-1)!!\,(N\langle\delta R^2\rangle)^{m/2}$ of \emph{all} even powers of
shifted nonzero eigenvalues in the limit of large $N$.
Interestingly, they agree with the central moments of a real Gaussian
distribution. The central moments of the eigenvalue density in the complex
plane are shown to satisfy exact sum rules involving these expectation values.

Correlations between eigenvalues are found to agree with the GinOE: Real
eigenvalues repel each other with an exponent of unity, complex eigenvalues are
repelled by the real axis with an exponent of unity and by each other with an
exponent of two.


In view of the power of RMT for Hamiltonians, we hope that
this approach will also benefit our understanding of complex stochastic
processes. Comparisons with real processes are now called for.


\appendix

\section{Large-$N$ limit for symmetric rate matrices}
\label{app.a}

In the limit of large $N$, the density of eigenvalues $\tilde\lambda$ of $\tilde
A$ only depends on the second moment $\langle \delta R^2\rangle$ of the
distribution of components $\tilde A_{ij}$, $i\neq j$, for any distribution
function of $\tilde A_{ij}$, as long as all its central moments exist.
In this appendix, we sketch the proof of this statement.


The eigenvalue density is given by Eq.~(\ref{2.rhodef}). In the expansion of the
geometric series for the resolvent \cite{SVW01},
\be
\langle \tilde G(z)\rangle = \sum_{n=0}^\infty
  \frac{\mathrm{Tr}\,\langle\tilde A^n\rangle}{z^{n+1}} ,
\label{app.a.dens}
\ee
the $n=0$ term is independent of the distribution of $\tilde
A_{ij}$, while the $n=1$ term vanishes. Since $\sum_i (\tilde A^n)_{ij}=0$ for
$n\ge 1$ we can write
\ba
\lefteqn{ \langle \tilde G(z)\rangle = \frac{1}{z} - \sum_{n=2}^\infty
  \frac{1}{z^{n+1}} \sum_{ij,i\neq j} \langle \tilde A^n \rangle_{ij} }
  \nonumber \\
& & = \frac{1}{z} - \sum_{n=2}^\infty
  \frac{1}{z^{n+1}} \sum_{ij,i \neq j} \sum_{k_1,k_2,\ldots} \!\langle \tilde
  A_{ik_1} \tilde A_{k_1k_2} \cdots \tilde A_{k_{n-1}j} \rangle .
  \nonumber \\
& & {}
\label{app.a.rhosum}
\ea
We now introduce a diagrammatic representation for the expectation values
$\langle\tilde A^m\rangle_{ij}$, $i\neq j$:
\be
\includegraphics[scale=0.5]{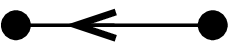} \;\equiv\;
  \! \sum_{ij,i\neq j} \langle\tilde A\rangle_{ij} = 0,
\ee
\be
\includegraphics[scale=0.5]{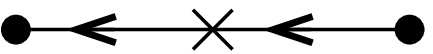} \;\equiv\;
  \! \sum_{ij,i\neq j} \! \langle \tilde A^2\rangle_{ij}
    = \sum_{ij,i\neq j} \sum_k \langle \tilde
  A_{ik}\tilde A_{kj}\rangle,
\ee
\be
\includegraphics[scale=0.5]{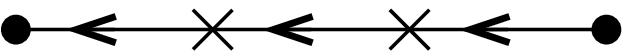} \;\equiv\;
  \! \sum_{ij,i\neq j} \langle \tilde A^3\rangle_{ij} \quad\mbox{etc.}
\ee
Here, an arrow represents a
factor of $\tilde A$, a vertex (filled circle or cross) represents a matrix
index, and all indices are summed over $1,\ldots,N$, subject to the
constraint that indices corresponding to filled circles are distinct.
Vertices drawn as crosses do not imply any constraint.


In Eq.~(\ref{app.a.rhosum}), we now decompose the sums over indices into
terms with equal and distinct indices. For equal indices we attach the
arrows to the same filled-circle vertex, whereas distinct indices are denoted
by distinct filled-circle vertices. For example,
\ba
\lefteqn{\sum_{ij,i\neq j} \langle \tilde A^2\rangle_{ij} =
  \;\parbox{2.4cm}{\includegraphics[scale=0.5]{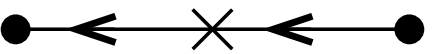}}} \nonumber \\
& & = \;\parbox{6cm}{\includegraphics[scale=0.5]{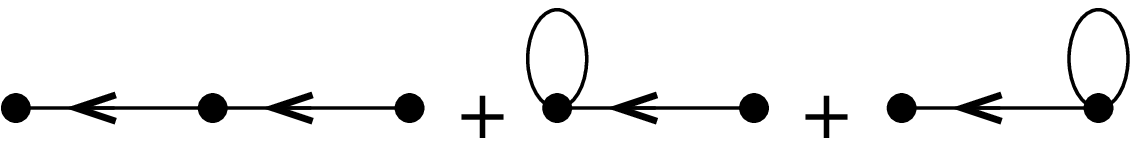}} .\qquad
\label{nof.ESRE.rules.a}
\ea
The constraint $\tilde A_{jj}=-\sum_{i\neq j} \tilde A_{ij}$ assumes the form
\be
\includegraphics[scale=0.5]{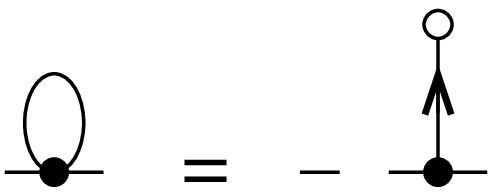} ,
\label{nof.ESRE.rules.b}
\ee
where the open circle denotes an index that is different from the one connected
to it but not otherwise constrained. Applying this rule to all
terms, we obtain open-circle vertices, which we dispose of by
again distinguishing between equal and distinct indices. For example,
\ba
\sum_{ij,i\neq j} \langle \tilde A^2\rangle_{ij} & = &
   \;\includegraphics[scale=0.5]{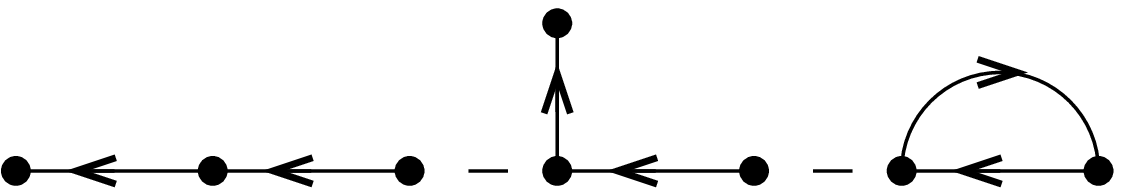} \nonumber \\
& & \includegraphics[scale=0.5]{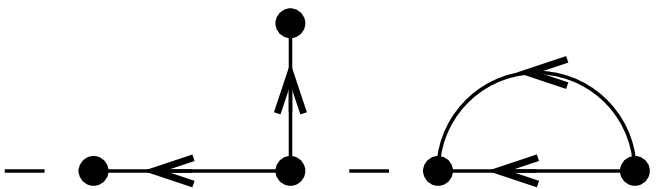} \;.
\label{nof.ESRE.rules.c}
\ea
We have achieved that factors of $\tilde A$ with two equal indices are no longer
present and that all indices to be summed over are distinct.


Since different off-diagonal components $\tilde A_{ij}$ are independent, except
for $\tilde A_{ji}=\tilde A_{ij}$, the expectation value of each term
decays into a product of expectation values of powers of components,
$\langle \delta R^m\rangle \equiv \langle (\tilde A_{ij})^m\rangle$. The
corresponding diagrams are of the forms
\ba
\includegraphics[scale=0.5]{diag1.eps} & = & 0 ,
\label{nof.ESRE.bubbles.a} \\
\parbox{3.2cm}{\includegraphics[scale=0.5]{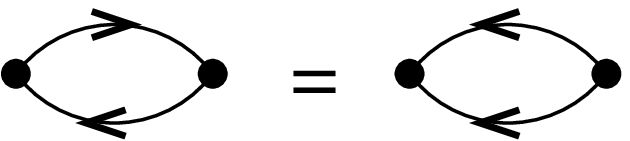}} & = &
  \langle \delta R^2\rangle ,
\label{nof.ESRE.bubbles.b} \\
\parbox{3.2cm}{\includegraphics[scale=0.5]{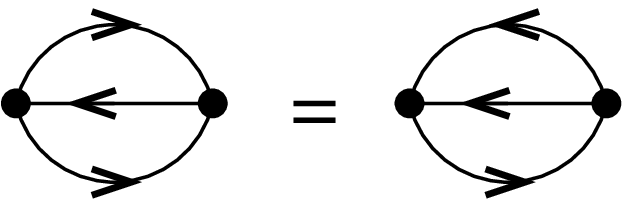}} & = &
  \langle \delta R^3\rangle \quad\mbox{etc.}
\label{nof.ESRE.bubbles.c}
\ea
Finally, any term containing $m$ vertices obtains a factor $N(N-1)(N-2)\cdots
(N-m+1)$ from the sum over distinct indices. In the limit of large $N$ this
becomes $N^m$.


We conclude that at any order $n\ge 2$ in Eq.~(\ref{app.a.rhosum}), the largest
terms for large $N$ are the non-vanishing ones with the maximum number of
vertices. Note that the diagrams generated by this procedure are always
connected. Diagrams containing single arrows connecting two vertices
vanish because of Eq.~(\ref{nof.ESRE.bubbles.a}). For even $n$, the maximum
number of vertices is $n/2+1$, which is
obtained if all connections are double arrows. In this case the contribution is
proportional to $N^{n/2+1}\langle \delta R^2\rangle^{n/2}$. The next smaller
terms have two triple arrows and contribute $\propto N^{n/2}\langle \delta
R^2\rangle^{n/2-3} \langle \delta R^3\rangle^2$. For odd $n$, the largest terms
have one triple
arrow and all other connections are double arrows. Their contribution is
proportional to
$N^{n/2+1/2}\langle \delta R^2\rangle^{n/2-3/2} \langle \delta R^3\rangle$.

Since Eq.~(\ref{2.rhodef}) contains an explicit factor of $1/N$,
the leading contributions to the density scale as $N^{n/2}$
($N^{n/2-1/2}$) for even (odd) $n$.
If we rescale the density so that the width approaches a constant, the
odd terms in the expansion (\ref{app.a.rhosum}) vanish like
$N^{-1/2}$, showing that the rescaled density approaches an even function.
Furthermore, the leading even terms only depend on the second
moment $\langle \delta R^2\rangle$, which is what we set out to prove.

Rewriting Eq.~(\ref{app.a.dens}) in terms of the moments $\mu_n$,
\be
\langle \tilde G(z)\rangle = \frac{1}{z} + (N-1) \sum_{n=2}^\infty
  \frac{\mu_n}{z^{n+1}} ,
\ee
we see that the terms of order $n$ contribute exclusively to the moment $\mu_n$.
The result proved here is consistent with the calculated
moments in Table \ref{tab.1}.



\section{Large-$N$ limit for general rate matrices}
\label{app.b}

For ensembles of general, asymmetric rate matrices, it is also true that
the density of eigenvalues only depends on the second moment $\langle \delta
R^2\rangle$ for large $N$. We here sketch the proof of this assertion.

The distribution of eigenvalues in the complex plane is given by
Eqs.~(\ref{3.Hdef})--(\ref{3.rhofullhermsd}). We define
\be
g(\eta;z,z^\ast) \equiv \mathrm{Tr}_{2N} \left(\begin{array}{cc}
  0 & I \\ 0 & 0
  \end{array}\right)
  \langle\mathcal{G}(0;z,z^\ast)\rangle
\ee
so that $\rho_\mathrm{all}(x,y) = (1/\pi N)\, \partial g(0;z,z^\ast)/\partial
z^\ast$ and expand the resolvent,
\ba
g(\eta;z,z^\ast) & = & \sum_{n\,\mathrm{odd}}
  \frac{1}{\eta^{n+1}}\,\mathrm{Tr}\,
  \Big\langle [(\tilde A^T-z^\ast I)(\tilde A-z I)]^{\frac{n-1}{2}} \nonumber
  \\
& & {}\times  (\tilde A^T-z^\ast I) \Big\rangle .
\label{app.b.g2}
\ea
Expanding the products, we obtain a linear
combination of expressions of the form $\mathrm{Tr}\,\langle \cdots \tilde A^T
\cdots \tilde A \cdots \rangle$ containing any number of factors $\tilde
A^T$ and $\tilde A$ in any order. Now the arguments of App.~\ref{app.a}
go through with few changes. We can group the terms
according to the total order $m$ of $\tilde A$ and $\tilde A^T$. The term of
order zero is independent of the distribution of $\tilde A_{ij}$. The terms of
first order are $\mathrm{Tr}\,\langle \tilde A\rangle = \mathrm{Tr}\,\langle
\tilde A^T\rangle = 0$. In all other terms we can use cyclic permutation under
the trace and the identity $\mathrm{Tr}\,B^T = \mathrm{Tr}\,B$ to make sure that
a factor $\tilde A$ and not $\tilde A^T$ is appearing first under the trace. We
can then use Eq.~(\ref{1.Aconstr}) to write
$\mathrm{Tr}\,\langle \tilde A \cdots\rangle = -\sum_{ij,i\neq j} \langle
\tilde A\cdots\rangle_{ij}$.

Now we can apply the diagrammatics of App.~\ref{app.a}. $(\tilde A^T)_{ij} =
\tilde A_{ji}$ is drawn as an arrow pointing in the opposite
direction. In the evaluation of expectation values corresponding to
Eqs.~(\ref{nof.ESRE.bubbles.b}), (\ref{nof.ESRE.bubbles.c}) we have to take into
account that $\tilde A_{ij}$ and $\tilde A_{ji}$ are now independent so that we
instead have
\ba
\includegraphics[scale=0.5]{diag1.eps} & = & 0 , \\
\parbox{1.15cm}{\includegraphics[scale=0.5]{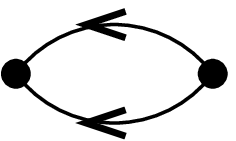}} & = & \langle \delta
  R^2\rangle ,
\label{nof.EGRE.bubbles.c} \\
\parbox{1.15cm}{\includegraphics[scale=0.5]{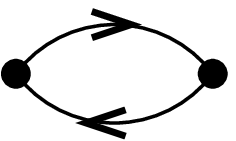}} & = & 0 , \\
\parbox{1.15cm}{\includegraphics[scale=0.5]{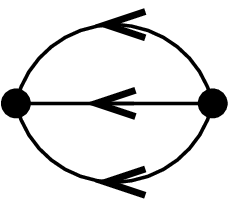}} & = & \langle
  \delta R^3\rangle , \\
\parbox{3.3cm}{\includegraphics[scale=0.5]{diag8.eps}} & = & \ldots = 0
\quad\mbox{etc.}
\ea
We note that all terms of the same order $m$ in Eq.~(\ref{app.b.g2})
have the same sign and thus cannot cancel. We thus find that to any order $m$
the leading terms in Eq.~(\ref{app.b.g2}) for large $N$ have the same form as
for symmetric matrices. In particular, for even $m$ the leading term in
the density $\rho(x,y)$ scales with
$N^{m/2} \langle \delta R^2\rangle^{m/2}$ and the odd terms scale with a lower
power of $N$. Finally, it is conceivable that
taking the derivative of $g(0;z,z^\ast)$ with respect to $z^\ast$ in order to
obtain the density could remove the leading-$N$ term. This is not
the case, since for any even order $m\ge 2$ there is at least a contribution
from $m=n-1$ in Eq.~(\ref{app.b.g2}), which is linear in $z^\ast$.

\section{Pseudomoments for the EGRE}
\label{app.c}

In this appendix, we use the diagrammatics of App.~\ref{app.a}
to calculate the pseudomoments
\be
\mu_m = \langle \tilde\lambda^m\rangle'
  = \frac{1}{N-1}\, \mathrm{Tr}\, \langle \tilde A^m\rangle ,
\ee
$m\ge 2$, to leading order for large $N$ for the EGRE. Appendix \ref{app.b}
shows that for large $N$ only the even pseudomoments are relevant. We write
\ba
\mu_m & = & -\frac{1}{N-1}\, \sum_{ij,i\neq j} \langle \tilde A^m\rangle_{ij}
\nonumber \\
& = & -\frac{1}{N-1}\, \sum_{ij,i\neq j} \sum_{k_1,k_2,\ldots} \!\! \langle
  \tilde A_{ik_1} \tilde A_{k_1k_2} \cdots \tilde A_{k_{m-1}j} \rangle .\qquad
\label{app.c.pseudom}
\ea
It was shown in App.~\ref{app.b} that for large $N$ the distribution of $\tilde
A_{ij}$ only enters through its second moment $\langle\delta R^2\rangle$.
We decompose all terms into a sum of contributions with equal or distinct
indices, see Eq.~(\ref{nof.ESRE.rules.a}). For each term, some or none of the
indices in $\{i,k_1,k_2,\ldots,k_{m-1},j\}$ are equal. Contributions for which
two equal indices are separated by other, distinct indices in this string,
correspond to diagrams of the type
\be
\parbox{1.15cm}{\includegraphics[scale=0.5]{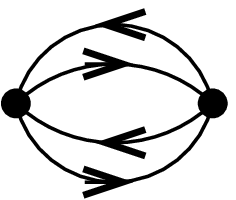}}
\ee
and are of lower order in $N$. All
remaining diagrams are of the form of chains leading from $j$ to $i$ with any
number of single-vertex loops ($\tilde A_{kk}$) decorating the vertices. We
call these single-vertex loops ``leaves.''

Next, we prove
\be
\parbox{6cm}{\includegraphics[scale=0.5]{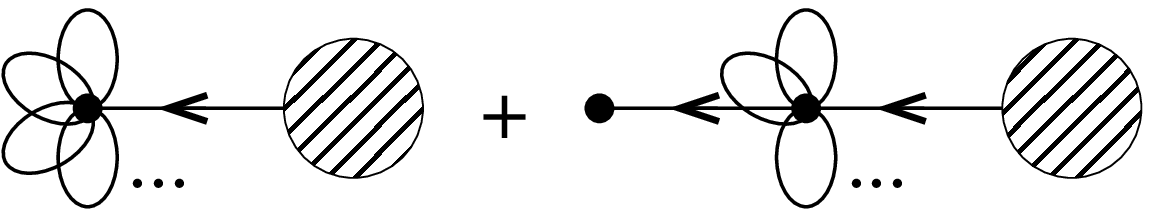}} \cong 0
\label{app.c.maincanc}
\ee
for $N\to\infty$, where the left-most vertex in the first term carries
$l\ge 1$ leaves, while the second vertex in the second term carries $l-1\ge 0$
leaves. The shaded circle is an arbitrary diagram part.
The proof proceeds as follows: Applying the rule
(\ref{nof.ESRE.rules.b}), we obtain
\be
\includegraphics[scale=0.5]{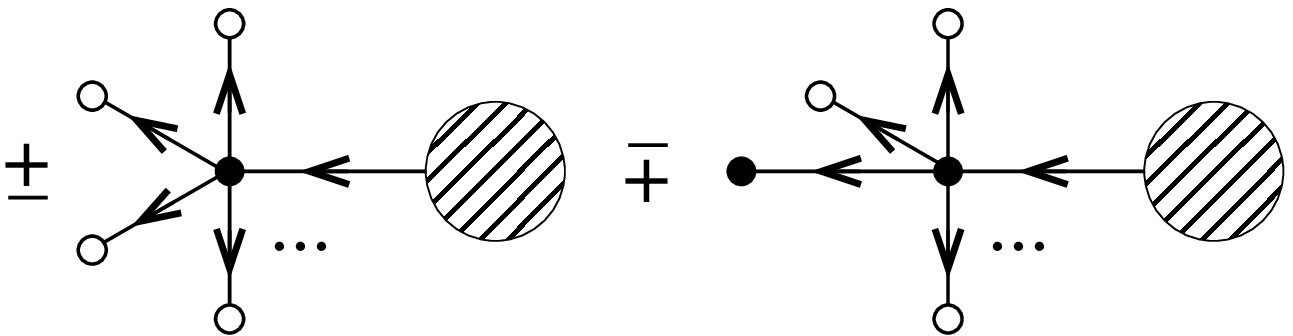}
\ee
with the upper (lower) signs for even (odd) $l$. In the leading large-$N$ term,
all connections must be of the form of two arrows pointing in the same
direction, as in Eq.~(\ref{nof.EGRE.bubbles.c}). This is only possible
if we pair up the open-circle vertices among themselves, not with any vertices
in the right-hand part of the diagrams. This requires $l$ to be even.
Furthermore, for the first diagram there are $(l-1)!!$ ways to partition $l$
leaves into pairs. For the second diagram there are $l-1$ ways to pair one of
the leaves with the leftmost vertex and $(l-3)!!$ ways to partition the
remaining $l-2$ leaves into pairs. With these factors we obtain
\ba
\lefteqn{(l-1)!!
\quad\parbox{2.5cm}{\includegraphics[scale=0.5]{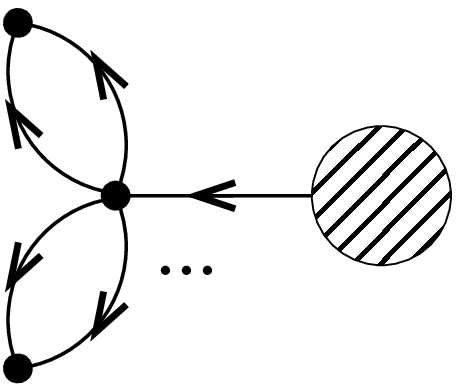}}} \nonumber
\\
& & {}- (l-1)(l-3)!!
\quad\parbox{3cm}{\includegraphics[scale=0.5]{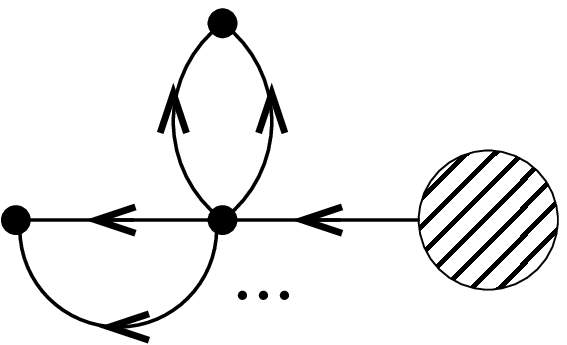}}
 = 0 .\quad
\ea
All diagrams of leading order in $N$ are of the form of one of the two diagrams
in Eq.~(\ref{app.c.maincanc}). Thus all diagrams cancel, except if only one of
the two forms exists. This is only the case for
\ba
\parbox{1.7cm}{\includegraphics[scale=0.5]{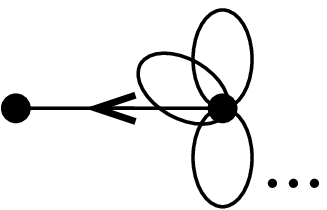}} & \cong & -(m-1)!!\,
\quad\parbox{1.7cm}{\includegraphics[scale=0.5]{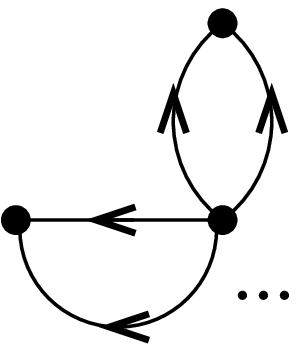}} \nonumber \\
& \cong & -(m-1)!!\, N^{m/2+1}\, \langle \delta  R^2\rangle^{m/2} ,\quad
\ea
since its partner would contain only a single vertex, which is
excluded by $i\neq j$. 


With the prefactor from Eq.~(\ref{app.c.pseudom}), we obtain
\be
\mu_m \cong (m-1)!!\, N^{m/2}\, \langle \delta R^2\rangle^{m/2} .
\label{app.c.pseudomres}
\ee



\end{document}